\documentclass[authoryear,12pt]{article}
\usepackage{graphicx}
\usepackage[authoryear]{natbib}
\usepackage{amsmath}
\usepackage{amssymb}
\usepackage{amsthm}
\usepackage{amsfonts}
\usepackage{color}
\usepackage{subfig}
\usepackage{float}

\newcommand{\field}[1]{\mathbb{#1}}

\begin{document}

\title{The multiplex structure of interbank networks\footnote{The views expressed in the article are those of the authors only and do not involve the responsibility of the Bank of Italy. The research leading to these results has received funding from the European Union, Seventh Framework Programme FP7/2007-2013  under grant agreement CRISIS-ICT-2011-288501 and from the INET-funded grant ``New tools in Credit Network Modeling with Heterogeneous Agents''. L. Infante contributed while visiting the NYU Department of Economics. All the usual disclaimers apply.}}
\author{L. Bargigli,$^\textrm{a}$
G. di Iasio,$^\textrm{b}$
L. Infante,$^\textrm{c}$
F. Lillo,$^\textrm{d}$
F. Pierobon$^\textrm{e}$}

\maketitle

\small
\begin{center}
$^\textrm{a}$~\emph{Scuola Normale Superiore, Pisa, Italy}\\
$^\textrm{b}$~\emph{Financial Stability, Bank of Italy}\\
$^\textrm{c}$~\emph{Economic Research and International Affairs, Bank of Italy}\\
$^\textrm{d}$~\emph{Scuola Normale Superiore di Pisa and Universit\`{a} di Palermo, Italy}\\
$^\textrm{e}$~\emph{Banking and Financial Supervision, Bank of Italy}
\end{center}
\normalsize

\begin{abstract}
The interbank market has a natural multiplex network representation. We employ a unique database of supervisory reports of Italian banks to the Banca d'Italia that includes all bilateral exposures broken down by maturity and by the secured and unsecured nature of the contract. We find that layers have different topological properties and persistence over time. The presence of a link in a layer is not a good predictor of the presence of the same link in other layers. Maximum entropy models reveal different unexpected substructures, such as network motifs, in different layers. Using the total interbank network or focusing on a specific layer as representative of the other layers provides a poor representation of interlinkages in the interbank market and could lead to biased estimation of systemic risk.
\end{abstract}


\color{black}
\section{Introduction\bigskip\protect}\label{intro}

In the wake of the recent financial crisis, interconnectedness among players in the financial system has become a major issue for regulators, as credit exposures among financial institutions have been one of the main vehicles of contagion during the crisis. The mapping of linkages among financial institutions has therefore become an essential tool in assessing systemic risk in the financial system.\bigskip

Network analysis has contributed to characterize, understand and model complex systems of interconnected financial institutions and markets \citep{gai2010contagion,battiston2012}. These tools are gaining popularity also among policymakers. Most contributions focus on the interbank market\footnote{See for instance \cite{boss2004network,soramaki2007topology,iori2008network,cont2011network,mistrulli2011assessing}.}, the plumbing of modern financial systems, especially in the Euro area. In the network perspective, the interbank market is commonly represented as a standard directed and weighted graph. Each link represents a credit relation between two counterparties. Directionality identifies the borrower and the lender; the weight of the link represents the loan amount. In general, the interbank market is much richer and complex than a simple weighted graph. In this paper we explore the differences in credit relations due to maturity of the contract or the presence of
collateral. Due to lack of data availability, existing empirical literature either (i) disregards the heterogeneity of credit relations or (ii) focuses only on one type, implicitly assuming that the network of the selected type of credit relations is a good proxy for the networks of other types. In the latter case, the vast majority of contributions focus on the overnight unsecured market. These two approaches are parsimonious but may provide biased results if the underlying ``representativeness'' assumptions fail.\bigskip

\begin{figure}[H]
\begin{center}
\includegraphics[width=0.7\textwidth,keepaspectratio=true]{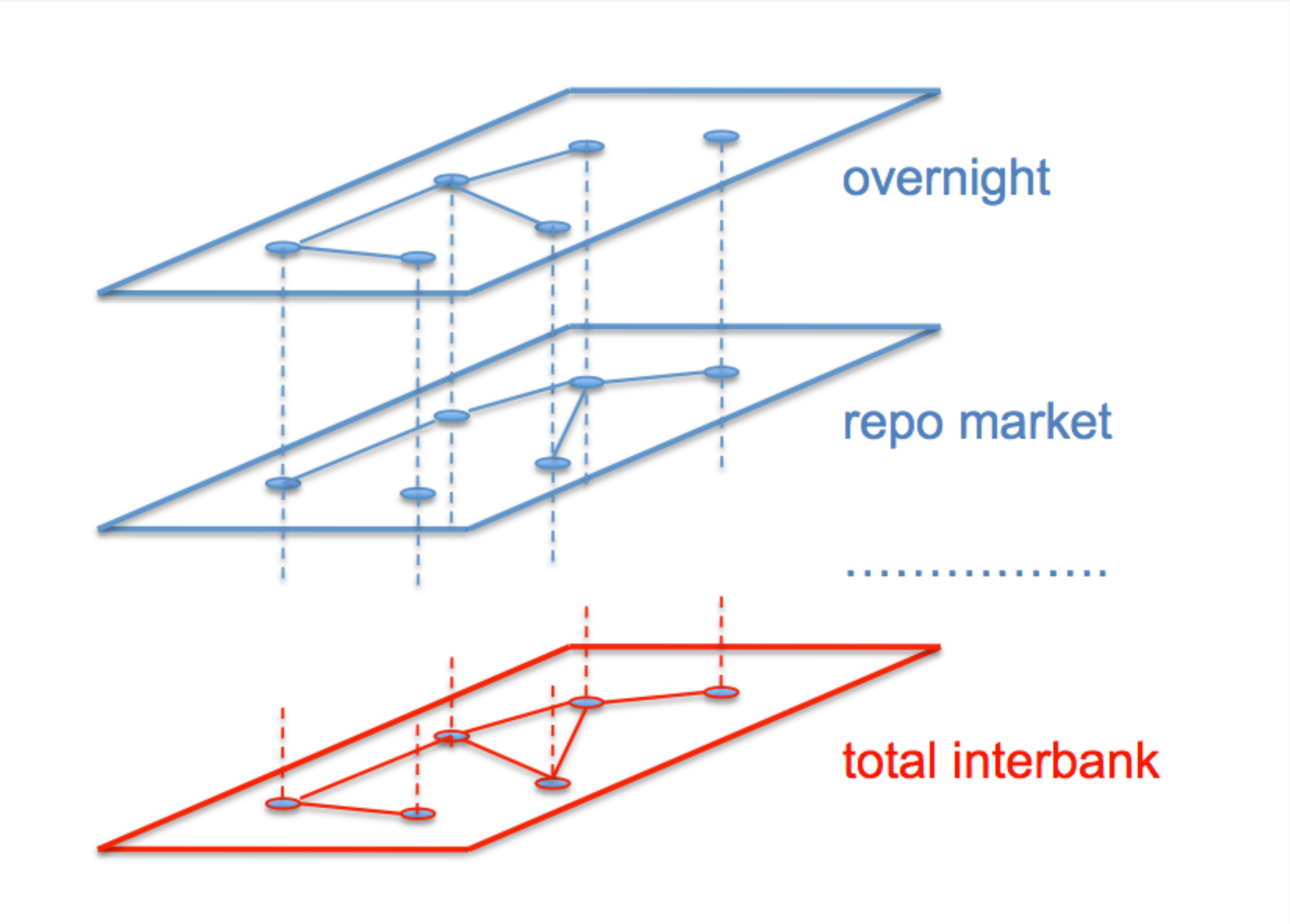}
\end{center}
\caption{\small{Stylized representation of the multiplex structure of the interbank market. Each node is a bank, and links represent credit relations. A layer (e.g. overnight, repo market,...) is the set of all credit relations of the same type. The network in red is the total interbank market, obtained by aggregating all the layers.}}
\label{multilayer}
\end{figure}

A much more accurate representation of the interbank market is a multiplex, or multilayer network. A multiplex (see Fig. \ref{multilayer}) is composed by a series of layers. Each node is a bank and each layer is a network representing one type of relations. The total (or aggregated) network is the aggregation of all the layers of the multiplex.

The general aim of this paper is to provide a broad analysis of differences and similarities between the layers of the interbank network. Our main research questions are: (1) Are the layers of the multiplex topologically different? (2) Is there a specific layer that leads the topological properties of the total network? (3) Is the occurrence of a link in a layer predictive of the occurrence of another link between the same nodes in a different layer or in the same layer at a different time?
\bigskip

We employ a unique database of supervisory reports of Italian banks to the Banca d'Italia. In brief, the dataset includes all bilateral exposures of all Italian banks, broken down by maturity and by the secured and unsecured nature of the contract. Moreover, we consolidate exposures at the banking group level and infra-group lending is netted out.\footnote{In what follows we provide evidence that the credit activity between banks belonging to the same banking group (internal capital market), dramatically differ in its characteristics and dynamics with respect to credit between banks belonging to different groups.}\bigskip

We perform three types of analysis to address our research questions. First, we compare the topological and metric properties of the different layers and of the total network. Similarity of topology, however, does not imply point-wise similarity. Therefore, the second set of analyses is aimed at quantifying the similarity between the different layers by computing the joint probability that a link between two nodes appears in more than one layer. Finally, we investigate the extent to which different random models inspired to the Maximum Entropy Principle fit with the layers of the multiplex.\bigskip
%

Our main results can be summarized as follows: (i) Different layers of the interbank network have several topological and metric properties which are layer-specific, while other properties seem to be more ``universal"; (ii) The topology of the total interbank market is closely mirrored by the one of the overnight market, while both are little informative about other layers; (iii) Higher order topological properties, such as triadic structures or reciprocated links, are explained by random models in some layers, but not in others; (iv) Random models that jointly consider topological and metric properties provide a close representation of the metric properties of different layers, as long as the calibration of the model is layer-specific.\bigskip

From a policy perspective, the heterogeneity of layers may be good news for financial stability, as it is likely to slow contagion across participants in the different layers of the market. The evidence that overnight unsecured layer most closely mirrors the topological features of the overall network should provide comfort to policymakers, as the overnight unsecured interbank market is the focus of monetary policy operations in several jurisdictions. Still, if policymakers and researchers alike were to target a specific segment of the interbank network, they should be careful in adopting an analytical framework based on the overall features of the network.\bigskip

{\it Relationship to the literature.} Multilayer network is a quite new branch of network theory and is becoming largely applied in many fields. While almost all network research has been focused on the properties of a single network that does not interact and depends on other networks, examples of interdependent networks abound in the real world \citep{gao2011robustness}. Many real-world networks interact with other networks. One may think of the internet, the airline routes and electric power grids. \cite{buldyrev2010catastrophic} find that interdependent networks become significantly more vulnerable to contagion compared to their non-interacting sub-networks.\bigskip

The economic literature on multilayer network is definitely not large. In general, our paper is related to several contributions that study the interbank market. Most of the studies on the interbank market focus on contagion. \cite{cont2011network} use a set of different kinds of inter-bank exposures (i.e. fixed-income instruments, derivatives, borrowing and lending) and study the potential contagion in the Brazilian market. Based on Italian data, \cite{mistrulli2011assessing} finds evidence that banksÕ default hardly triggers a systemic crisis. \cite{iazzetta2009topology} focus on Italian bilateral interbank deposits, in order to investigate  the centrality of the banks and the resiliency of the system. Other studies contributed to this line of research making use of estimation techniques to reconstruct bilateral relationships (see \cite{upper2011simulation} for a complete survey).
\cite{montagna2013multi} develop an agent-based model with the aim to catch risks arising from different banksÕ businesses. In this case the network representation of the interbank market relies on a multi-layered model that takes into account long- and short-term bilateral exposures and common exposures to external financial assets. The authors find that the interactions among the layers matter in amplifying the contagion risk. As a byproduct, focusing on a single interbank segment can underestimate shocks propagation. \cite{abbassi2013} use an estimated dataset of interbank exposures from Target 2 payment system and embed network covariates into their econometric model. They estimate the impact of the Lehman collapse on the Euro interbank market structure. Market segments with different maturity have reacted in different ways. No decline in volume is recorded in the overnight unsecured segment, while the term money markets dropped after the Lehman bankruptcy.\bigskip
 
Many other non-network papers attempt to evaluate the effects of the financial crisis in different interbank market segments. \cite{afonso2010} analyze increased counterparty risk and liquidity hoarding in the US federal fund market after the Lehman episode. \cite{kuo2013} study the US term interbank market, using both quantity and price information. The main finding of the paper is a shortening in the maturity along with a reduction in the volume exchanged around the Lehman crisis.\bigskip

The paper is organized as follows. Section \ref{data} presents our dataset. Section \ref{results} is devoted to the comparison of the topological and metric properties of the different layers. In Section \ref{sec:similarity} we present a similarity analysis. In Section \ref{sec: null_models} we investigate how statistical ensembles of networks are able to explain the structure of the different layers. Three technical appendices present the mathematical tools used in our analysis.

\section{Data description}\label{data}

We build a dataset of interbank transactions based on the supervisory reports transmitted to Banca d'Italia by all institutions operating in Italy. Supervisory information covers both locally incorporated banks and Italian branches of foreign banks. We focus on information reported by lending institutions at the time of the mapping of the network structure. Banks also report transactions with foreign institutions. However, this information cannot be fully employed in assessing the characteristics of the interbank network for several reasons. Then, we focus only on domestic links.\bigskip 

Information refer to end-of-year outstanding balances. We focus on 5 observations starting from end 2008. Over this period, Italian banks experienced the dry up of international interbank market that followed the September 2008 Lehman crisis as well as the impact of the Eurozone sovereign debt crisis which, particularly since the fall of 2011, sharply reduced the amount of interbank funds made available to Italian institutions due to sovereign risk concerns by market participants. The response of the Eurosystem consisted of liquidity injections, like other central banks worldwide, to alleviate tensions in the money market.  Extraordinary measures were introduced.\footnote{The most important are the fixed rate full allotment procedure, the supplementary longer-term operations, the broadening of eligible collateral, two Covered Bond Purchase Programmes and currency swaps with a number of major central banks. In addition, two longer-term refinancing operations (LTROs) with 36-month maturity were allotted on 21 
December 2011 and 29 February 2012, with a temporary widening of the collateral framework.} Moreover, a wave of regulatory interventions and proposals are affecting bank's funding structure and the interbank market. As a major European country, Italy and its interbank market have experienced this sequence of shocks and policy interventions and represent a quite unique environment.\bigskip

Most banks operate in Italy through a large set of subsidiaries as a result of the consolidation process that took place in the 1990's and the 2000s: for example, the 5 largest groups - which currently account for some 60 per cent of the total banking assets - operate through some 55 subsidiaries. Distinguishing between intragroup and intergroup transactions is therefore crucial. Since interbank lending and borrowing decisions are normally taken at the parent company level, we assume that the relevant economic agents of the network lie at the group level. We thus focus on data on intergroup transactions consolidated at the group level. Still, internal capital markets play a large role in the network, as funds managed by the parent company are distributed among the group's subsidiaries. We retain information on intragroup contracts by modeling such transactions as self-loops, in which the lending and borrowing institutions coincide. Network statistics are adjusted by dropping such relationships whenever
it is required by the definition of the corresponding metric.\bigskip

The representation of the interbank market is a weighted and directed network, namely a set of nodes (banks) that are linked to each other through different types of financial instruments (edges). The direction of the link goes from the bank $i$ having a claim to the bank $j$, and the weight is the amount (in millions of euros) of liabilities of $j$ towards $i$.\bigskip

The availability of other information on interbank transactions shapes the nature of our multi-layer analysis in which we focus on the breakdown of the network by contract type and maturity. The most relevant types of transactions recorded include overnight, sight and term deposits, certificate of deposits and repurchase agreements. For the sake of simplicity we only distinguish between unsecured and secured transactions, while we retain information on the maturity of unsecured contracts by distinguishing between overnight, short term (up to 12 months excluding overnight) and long term transactions (more than 12 months). We remark that our data on secured transactions only refer to OTC contracts, while the vast majority of secured transactions take place on regulated markets and are centrally cleared.\bigskip

The Italian financial system is bank-based and the Italian Interbank Network (IIN) is one of the largest in the Euro area. Table \ref{tab:volumes} reports end-of-period outstanding amounts. The top panel refers to non-consolidated data that include intragroup lending. Consolidated data, netted out of intragroup transactions are reported in the bottom panel. A very large fraction of the interbank network is due to intragroup lending (more than $80\%$ on average); moreover, a big drop in intragroup volumes is reported in 2010, due to the merging of several subsidiaries of a major group. The observed large differences between the two panels of Table \ref{tab:volumes} confirm that non consolidated data would offer a blurred picture of the network of market transactions.\bigskip 

\begin{table}[h]
 \centering
 \subfloat[Non consolidated data (intragroup lending is included)]{
\begin{tabular}{lrrrrr}
\hline
Layer & 2008 & 2009 & 2010 & 2011 & 2012 \\
\hline
Unsecured overnight &185 &147 &71 &68 &79\\
Unsecured ST &157 &192 &97 &97 &81\\
Unsecured  LT &68 &110 &95 &102 &103\\
Secured ST &74 &39 &43 &65 &36\\
Secured  LT &0.1 &8.0 &0.8 &2.5 &4.9\\
\hline
Total &485 &497 &308 &336 &306\\
\hline
\end{tabular}} \\
\subfloat[Consolidated data (intragroup lending is excluded)]{
\begin{tabular}{lrrrrr}
\hline
Layer & 2008 & 2009 & 2010 & 2011 & 2012 \\
\hline
Unsecured overnight &22 &19 &16 &17 &19\\
Unsecured ST &26 &27 &13&14 &12\\
Unsecured  LT &6 &3 &7 &17 &28\\
Secured ST &15 &5 &17 &11 &6\\
Secured  LT &0.03 &0.3 &0.7 &0.6 &1,4\\
\hline
Total &70 &55 &54 &61 &68\\
\hline
\end{tabular}}
 \caption{\small{Domestic credit exposures in the Italian interbank market. Billions of euros. End-of-period outstanding amounts.}}
\label{tab:volumes}
\end{table}

Looking at consolidated data, after a decline of $25\%$ from 2008 to 2010, the 2012 figures are close to those of 2008. The overnight interbank market, quite often at the core of network analysis, is responsible for roughly one third of the total volume. The time dynamics of the outstanding amounts in different layers are very heterogeneous, 
In particular, from 2010 we observe a decline in the unsecured short-term layer, mirrored by an increase of unsecured long-term lending. The surge of long-term interbank activities is consistent with incentives embedded in the recently established requirements on liquidity risk. Indeed, the Basel Liquidity Coverage Ratio significantly encourages borrowers to lengthen the maturity of their liabilities. The push towards long-term maturities also mirrors the shift of interbank lending from a money market, risk-free activity to a credit-intensive one, as a result of repeated bank failures and seizures during the financial crisis.\bigskip

The drop in secured interbank lending is mostly due to regulatory incentives towards centrally cleared as opposed to over-the-counter repurchase agreements, which turns into increasing costs of establishing bilateral lending agreements. Nevertheless, a number of banks are keen on keeping trading on a bilateral basis for the purpose of diversifying their funding sources.

%

\section{The Multiplex Italian interbank market}\label{results}

The topological and metric properties of the Italian Interbank Network (IIN) in the 2008-2012 period are investigated by comparing the properties of the total network - resulting from the aggregation of all the layers of the multiplex - with those of individual layers. Our time interval covers the early stage of the financial crisis, the Euro zone sovereign debt crisis and the following ECB extraordinary interventions.\bigskip

We start by providing a comprehensive set of network metrics for the total network and for each layer (see Appendix A for the definitions of the metrics). Table \ref{tab:summarynet} summarizes some results. Existing empirical works \citep{boss2004network,iori2008network,cont2011network,RePEc:kie:kieliw:1759,RePEc:fip:fednsr:354} detect some statistical regularities in interbank networks: (i) sparsity and low average distance between nodes, (ii) heterogeneity of nodes' degree, often associated with a power law tail distribution of degree with a small exponent ($< 3$), (iii) disassortative mixing, i.e. the tendency of high degree nodes to connect with low degree nodes, (iv) small clustering, and (v) heterogeneous level of reciprocity. As mentioned above, these studies typically focus on the overnight market, as data are more easily available. Alternatively, some contributions analyze the total network, in the aggregate, without a detailed analysis at the layer level. According to our first very simple analysis 
that follows, the total network and the overnight layers share very similar topological properties;
 conversely, other layers have significant differences in topology. A general implication is that systemic risk assessment (e.g. contagion analyses) based on the total interbank market (i.e. missing granularity of different layers) would exclusively reflect the properties of the overnight segment. However, the latter is a very poor approximation of other layers, which may be non-negligible in terms of size and, potentially, for systemic risk assessment.\bigskip

More in detail, the metrics of the total network are quite stable over the 2008-2012 period. The number of banking groups slightly declines from $573$ in 2008 to $533$ in 2012. The network is very sparse (the density is  approximately 1\%), weakly completely connected and almost strongly completely connected. The average path length is low, indicating a compact network structure. Regarding individual layers, the figures of the overnight segment are very close to those of the total network. Almost all banking groups that operate in the total network also trade in the overnight market. This layer is almost connected, both weakly and strongly. Similarly, the unsecured short-term layer involves almost all banking groups, but it is characterized by a much lower density.
The increase in volume of the unsecured long-term layer (see Table \ref{tab:volumes}) is clearly associated with the rising number of banks operating in this layer (from 238 to 450). It is much less dense than the overnight layer. The strong component is far smaller than the weak component. The two secured layers are much smaller in size as nowadays the majority of collateralized trades are operated through Central Counterparties (CCPs). In particular, the secured long-term layer is a very small network. The secured short-term layer has less than 100 nodes.


\begin{table}[t]
\begin{tabular}{|l|rrrrrr|}
\hline
Statistics (2008) & U OVN & U ST & U LT & S ST & S LT &  TOT\\
\hline
\# of nodes &  573 &  550 &  238 &  72 &  8 &  573\\
\# of edges  &  2936  &  1457  &  354 &  125 &  7 &  3534 \\
Density &  0.8\% &  0.5\% &  0.6\% &  2.4\% &  12.5\% &  1.0\% \\
Largest weak compon.&  573 &  549 &  230 &  48 &  6 &  573\\
Largest strong compon.&  498 &  333 &  27 &  14 &  1 &  528\\
Avg undir. path length &  2.3 &  2.5 &  3.1  &  2.3&  1.8  &  2.2\\
Avg dir. path length &  2.4 &  2.7 &  2.4 &  1.9 &  - &  2.3\\
\hline
\end{tabular}
\begin{tabular}{|l|rrrrrr|}
\hline
Statistics (2012) & U OVN & U ST & U LT & S ST & S LT & TOT\\
\hline
\# of nodes &  532 &  521 &  450 &  45 &  18 &  533\\
\# of edges &  2560 &  1254 &  887 &  67 &  25&  3235\\
Density &  0.8\%  &  0.4\%  &  0.4\%  &  3.3\%  &  7.9\%  &  1.0\% \\
Largest weak compon.&  532  &  520  &  447  &  35  &  11  &  533 \\
Largest strong compon.&  456 &  375  &  165  &  16  &  3  &  513 \\
Avg undir. path length &  2.3  &  2.6 &  2.6  &  2.7  &  1.7 &  2.2 \\
Avg dir. path length &  2.4  &  2.8  &  2.8 &  2.5 &  1.3  &  2.4\\
\hline
\end{tabular}
\caption{\small{Summary statistics of network metrics of the aggregated IIN and its layers in 2008 and 2012. U = Unsecured ; S = Secured; OVN = Overnight;  ST = maturity less than one year (in the unsecured segment, the overnight is excluded);  LT = maturity more than one year.}}
\label{tab:summarynet}
\end{table}

\bigskip

\begin{figure}[H]
\begin{center}
\subfloat[In-degree, 2012]{\includegraphics[width=0.5\textwidth,keepaspectratio=true]{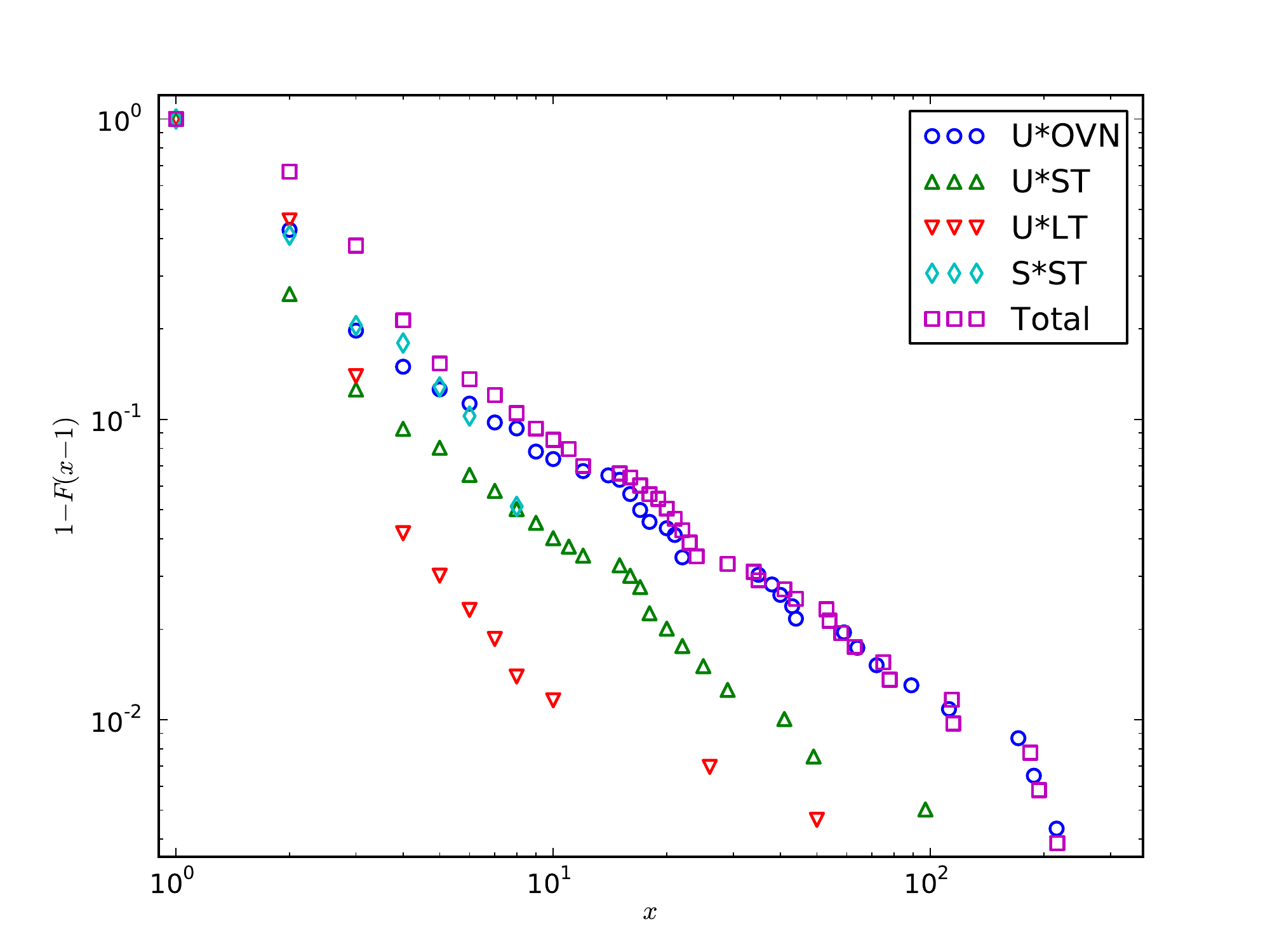}}
\subfloat[Out-degree, 2012]{\includegraphics[width=0.5\textwidth,keepaspectratio=true]{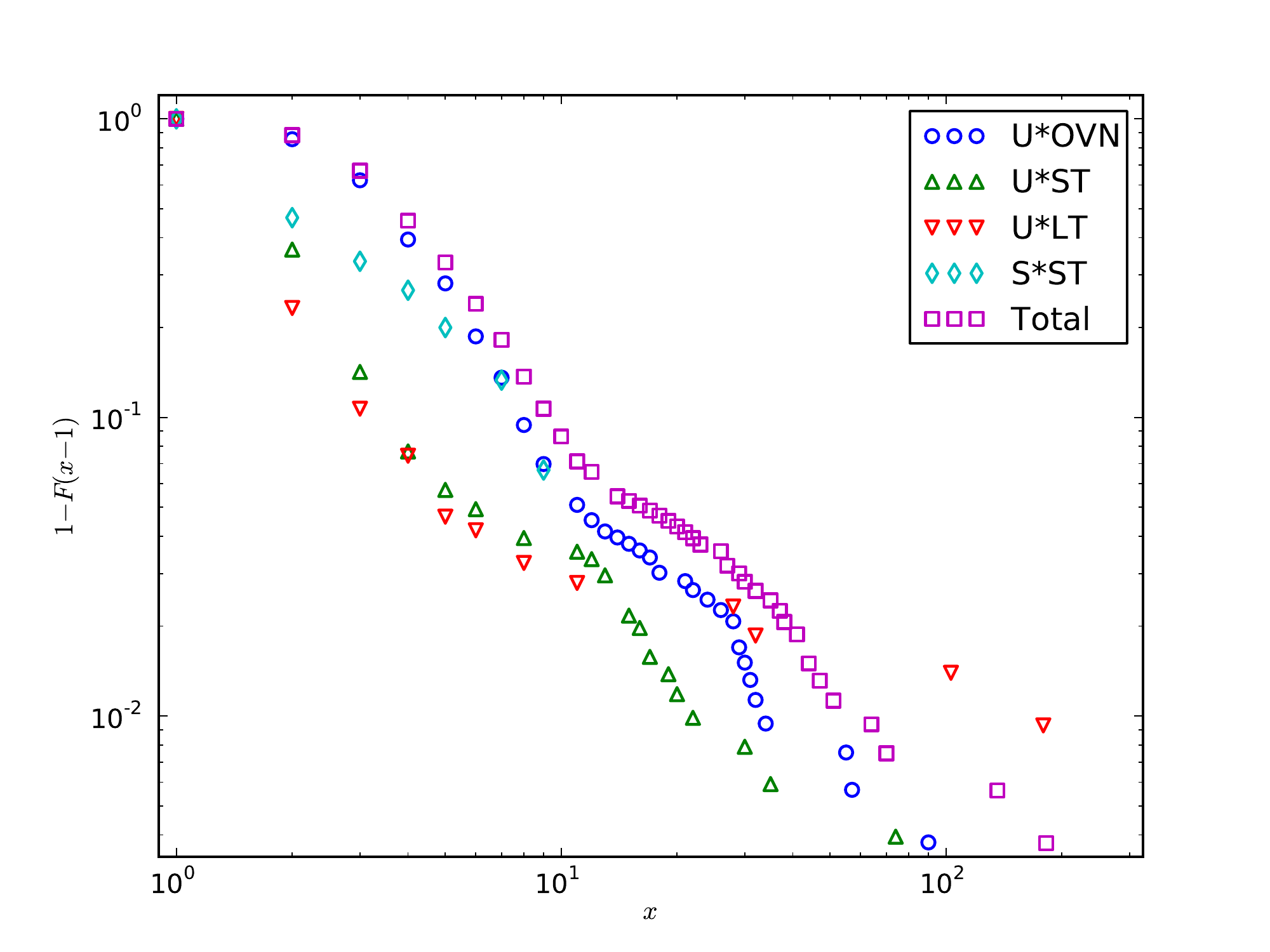}}
\end{center}
\caption{\small{Complementary cumulative distribution function of in-degree (left panel) and out-degree (right panel) of the total IIN and of its layers. The plots are in log-log scale.}}
\label{degree}
\end{figure}

One key stylized fact observed on real data is that interbank networks have a scale free structure. Technically, scale-free networks are characterized by a complementary cumulative distribution function (CCDF) of degree (or weight) which is asymptotically described by a power law functional relation.\footnote{A power law function is $P(X>x)\sim x^{-\alpha}$, where $\alpha$ is the tail exponent.} The peculiarity of scale-free interbank networks is the relative abundance of banking groups whose degree significantly exceeds the average: these ``hubs'' explain a large portion of lending transactions. Figure \ref{degree} shows the log-log plot of CCDF of in-degree and out-degree of the different subnetworks of the IIN in 2012.

We find (i) a fat tail (i.e. non exponential) asymptotic behavior for total network and for all layers; (ii) the tail exponent is similar across layers; (iii) the degree distributions of the total and of the overnight networks are very similar (especially for in-degree). We estimate the tail exponent $\alpha$ by using the maximum likelihood method developed by \cite{clauset2009power}, which identifies both $\alpha$ and the power law region. The estimated values of $\alpha$ are remarkably stable across layers and over time. The range of values of $\alpha$ is $[1.8,\,3.5]$ with most values concentrated around $\alpha= 2.3$. We also perform a log-likelihood ratio test of  power-law against lognormal. In the vast majority of cases (layers and years), the power law hypothesis cannot be rejected. Exceptions (lognormal preferred to power law) are observed for almost all years in the out degree of the secured short term layer.\bigskip

The Spearman correlation between degree and strength is used to test whether very interconnected banks have also large size credit relations (Table \ref{tab: spearman}). In all cases, a high and statistically significant correlation is obtained. In the overnight market the correlation is much lower than in the other layers: almost all banks operate in this segment which, in turn, drives the result of the total network.\bigskip

\begin{table}[H]
\centering
\subfloat[out-degree vs out-strength]
{
\begin{tabular}{l|ccccc}
\hline
Layer & 2008 & 2009 & 2010 & 2011 & 2012\\
\hline
U OVN & 0.4433 & 0.4755 & 0.4680 & 0.4933 & 0.5190\\
U ST & 0.5974 & 0.6313 & 0.6300 & 0.5596 & 0.5906\\
U LT & 0.9372 & 0.9172 & 0.9430 & 0.9530 & 0.9391\\
S ST & 0.9119 & 0.9579 & 0.9972 & 0.9913 & 0.9339\\
Total & 0.5564 & 0.5641 & 0.5355 & 0.5115 & 0.5192\\
\hline
\end{tabular}
} \\
\subfloat[in-degree vs in-strength]
{
\begin{tabular}{l|ccccc}
\hline
Layer & 2008 & 2009 & 2010 & 2011 & 2012\\
\hline
U OVN & 0.7332 & 0.7756 & 0.7050 & 0.6423 & 0.7081\\
U ST & 0.9018 & 0.9009 & 0.9008 & 0.8053 & 0.7545\\
U LT& 0.9134 & 0.8776 & 0.7373 & 0.4695 & 0.4939\\
S ST & 0.9508 & 0.9115 & 0.6128 & 0.7816 & 0.8084\\
Tot & 0.7562 & 0.7414 & 0.6972 & 0.5612 & 0.5066\\
\hline
\end{tabular}}
\caption{\small{Spearman correlation coefficient between degree and strength (in and out) of the nodes in the different layers of the IIN.}}
\label{tab: spearman}
\end{table}

Correlation is stronger in the (OTC) secured segment: the number of institutions active in this market is quite limited compared to other layers, as fixed costs of establishing bilateral lending agreements - such as ICMA Global Master Repurchase Agreement - favor large over small transactions and therefore big over small players. A similar case can be made for unsecured long-term transactions, whereas in short-term unsecured markets big players with high degree might exchange relatively small quantities with a large number of small banks, thereby pushing correlation coefficients down.\bigskip

Unlike many other socio-economic systems, interbank networks are observed to be disassortative. A network is disassortative when high degree (or weight) nodes tend to be connected to other high degree (or weight) nodes less frequently than expected under the assumption of a random rewiring of the network that preserves each node's degree (or weight). Table \ref{tab:assortative} reports the in- and out- assortativity in 2008 and 2012, considering both degree and weight. 

\begin{table}[H]
\begin{tabular}{|l|rrrrrr|}
\hline
Date: 2008 & U OVN & U ST & U LT & S ST & S LT & TOT\\
\hline
Out-degree assort.&  -0.26** &  -0.40** &  -0.52** &  -0.43** &  0.00 &  -0.27** \\
In-degree assort&  -0.34** &  -0.32** &  -0.35** &  -0.32** &  0.15 &  -0.33**\\
Out-weight assort.&  -0.02 &  -0.01 &  -0.06 &  -0.18* &  -0.21 &  -0.05**\\
In-weight assort.&  -0.03* &  -0.01 &  -0.06 &  -0.14 &  0.38 &  -0.06** \\
\hline
Degree reciprocity &  0.43* &  0.45* &  0.10* &  0.18* &  0.14 &  0.47* \\
Weight reciprocity &  0.43* &  0.16* &  0.05* &  0.13* &  0.04 &  0.29* \\
\hline
\end{tabular}\\
\begin{tabular}{|l|rrrrrr|}
\hline
Date: 2012 & U OVN & U ST & U LT & S ST & S LT & TOT\\
\hline
Out-degree assort.&  -0.27** &  -0.40** &  -0.51
** &  -0.17 &  0.06 &  -0.31**\\
In-degree assort.&  -0.42** &  -0.39** &  -0.38** &  -0.31* &  0.12 &  -0.37**\\
Out-weight assort.&  -0.03 &  -0.05 &  -0.32** &  -0.16 &  -0.29 &  -0.11** \\
In-weight assort.&  -0.18** &  -0.04 &  -0.03 &  -0.15 &  -0.05 &  -0.07** \\
\hline
Degree reciprocity &  0.40** &  0.56** &  0.31** &  0.31** &  0.05 &  0.45** \\
Weight reciprocity &  0.20** &  0.00** &  0.01** &  0.05* &  -0.00 &  0.07** \\
\hline
\end{tabular}
\caption{\small{In- and out-degree and weight assortativity of the Italian Interbank Network in 2008 (top table) and 2012 (bottom table). The lower part of each table shows the degree and weight reciprocity. One asterisk (two asterisks) denotes statistical significance at 5\% (1\%) confidence level.}}
\label{tab:assortative}
\end{table}

\begin{figure}[H]
\begin{center}
\includegraphics[width=10cm,height=6cm]{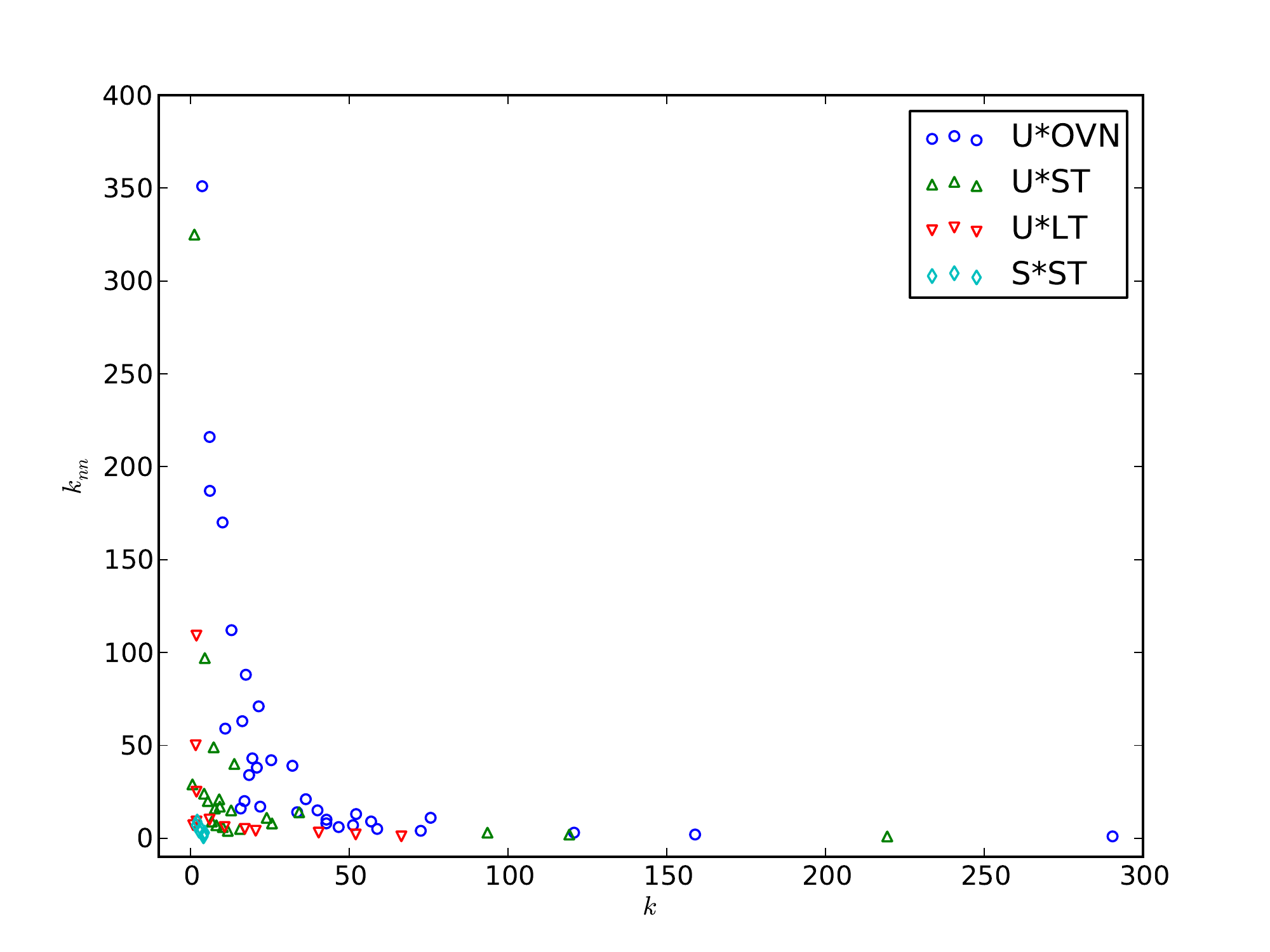}
\end{center}
\caption{\small{Relation between the in-degree (x-axis) of a node and the average degree of its neighbors (y-axis) in 2012. The results for the total network are very similar to the one for the overnight market and are not shown for the sake of clarity.
}}
\label{clustering1}
\end{figure}

Figure \ref{clustering1} presents the average in-degree of neighbors of a bank as a function of its in-degree. The monotonically decaying behavior is a clear sign of disassortativity. All layers have similar assortativity properties.

Table \ref{tab:assortative} also reports the results of the analysis of the reciprocity (degree and weight) of the different layers of the interbank network. In general, reciprocity is significantly higher in the overnight layer, especially considering weights. The only exception is the degree reciprocity of the unsecured short-term layer, which is also pretty high. However when one considers weights, reciprocity drops significantly as compared with the weight reciprocity in the overnight market. In general, many banks still hold bilateral deposit accounts at other banks for the settlement of retail payments. These accounts may be reported as overnight lending transactions even though they do not stem from a credit decision by the bank. Reciprocity in the unsecured overnight layer maybe therefore overestimated as a result.\bigskip

Clustering is the tendency of the neighbors of a node to connect with each other. Table \ref{tab:clustering} shows the average directed and undirected clustering of the different layers. The coefficients for the total network are surprisingly high. It is worth noting that most of the clustering is due to the overnight segment, while other layers display much lower values. Even in this case, clustering properties of the interbank market based on the overnight market (or on the total interbank market) offer a partial picture of other layers.

\begin{table}[H]
\begin{tabular}{|l|llllll|}
\hline
Date: 2008& U OVN & U ST & U LT & S ST & S LT & TOT\\
\hline
Avg dir. clustering &  0.393 &  0.112 &  0.056 &  0.161 &  0.135 &  0.463 \\
Avg undir. clustering &  0.527 &  0.170 &  0.083 &  0.180 &  0.270 &  0.571\\
\hline
\end{tabular} \\
\begin{tabular}{|l|llllll|}
\hline
Date: 2012& UOVN & U ST & U LT & S ST & S LT & TOT\\
\hline
Avg dir. clustering &  0.402 &  0.131 &  0.156 &  0.118 &  0.184 &  0.448\\
Avg undir. clustering &  0.547 &  0.209 &  0.303 &  0.169 &  0.311 &  0.577\\
\hline
\end{tabular}
\caption{\small{Clustering coefficient (directed and undirected) of the different layers of the IIN in 2008 and 2012.}}
\label{tab:clustering}
\end{table}
 
\begin{figure}[H]
\begin{center}
\includegraphics[width=10cm,height=6cm]{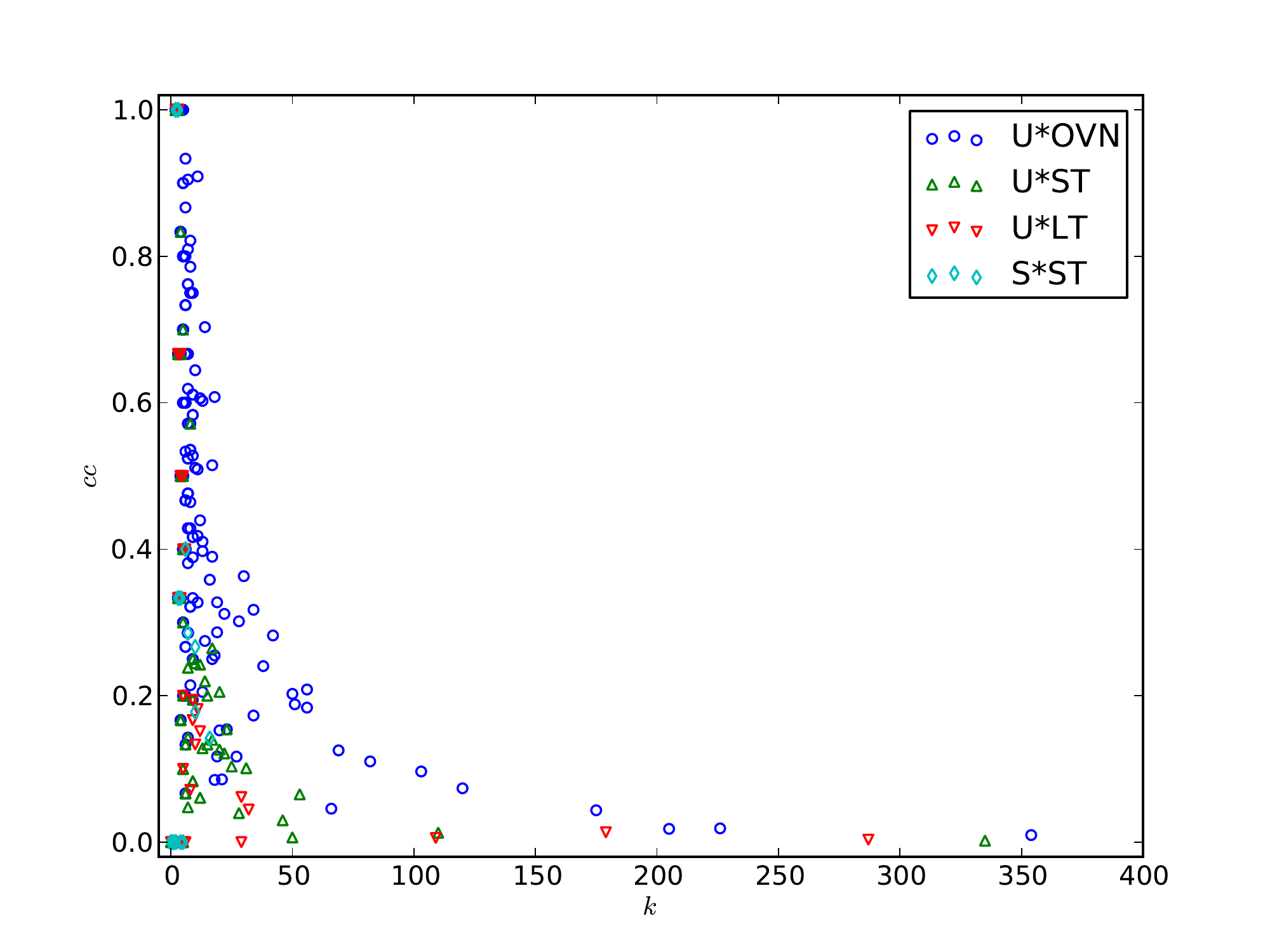}
\end{center}
\caption{\small{Scatter plot of the clustering coefficient of a node versus its degree in 2012. The results for the total network are very similar to the one for the overnight market and are not shown for the sake of clarity.
}}
\label{clustering2}
\end{figure}

Quite commonly, there is an inverse relationship between clustering and degree. Figure \ref{clustering2} shows the scatter plot of the two coefficients for each layer. The relatively high level of clustering is likely to be driven by the use of consolidated data, casting some doubt on the standard result of low clustering as a fundamental property of interbank networks. Moreover, given the inverse relationship between degree and clustering, the average value of clustering is mostly determined by the contribution of low degree nodes. Thus the resulting value is determined, to a large extent, by the degree distribution of the network. In order to correct this bias, in section \ref{sec: null_models} we focus on the number of triangles (instead of clustering coefficients), obtaining very different results.

The clustering coefficient of hubs is very close to zero in most layers, suggesting that hubs behave essentially as star centers in the network for all layers other than the overnight one, through which hubs connect their neighbors to other market participants.

\section{Results from similarity analysis}\label{sec:similarity}

In this section we focus on the stability of individual layers over time and  the similarity between different layers, at a certain point in time. In general, two networks can have very similar topological properties but the existence of a link between two nodes in one network may give no information on the probability that the same two nodes are linked in the other network as well. When the two networks are the realizations of the same layer at two different instant of time, the similarity between the two is a measure of the stability over time of the network structure of the layer. When the two networks of the previous example are two layers of a multiplex, the similarity analysis assesses to what extent a layer is representative of the other. Similarity analysis is a relevant tool in assessing financial stability: diffusion properties - and therefore contagion - in a multiplex depends on the similarity between the layers \citep{Gomez13}.\bigskip

There are many ways to define network similarity. Appendix \ref{appendixb} reviews shortly the main concepts put forward by the literature. All measures represent a network as an ordered vector. We use the Jaccard similarity $J$, defined by eq. (\ref{eq: jaccard}), for binary networks. Jaccard similarity can be interpreted as the probability of observing a link in a network conditional on the observation of the same link in the other network. Cosine similarity, defined by eq. \ref{eq: cosine}, is used instead for weighted networks.\bigskip

Two layers must share the same set of nodes in order to perform similarity analysis between them. In the following analysis, we either include only the nodes which enter in both layers (i.e. we take the intersection of the two node sets) or we include all nodes which enter in at least one layer (i.e. we take the union of the two node sets).\bigskip

Our results show that different layers display very different time persistence. The topology of the overnight layer appears to be more stable, with $J$ values roughly around 70\% between two successive years. The measure $J$ progressively declines increasing the lag. The topological similarity is not substantially affected by focusing only on common nodes (intersection instead of union): the set of nodes that operate in this market is relatively stable  over time. The unsecured short-term layer is slightly less persistent ($J \approx 60\%$) and still stable restricting to the intersection. Conversely, the unsecured long-term layer is highly variable, especially in the union, with $J = 70\%$ between 2008-2009 and between 2011-2012, and a much lower similarity ($J \approx 40\%$) between the other consecutive years. Finally, the secured short-term layer is quite volatile, both in terms of operating nodes and in terms of links. In fact, the similarity between consecutive years drops from 40-50\%
 to 20-30\% when moving from intersection to union. Unsurprisingly, similarity is lower when weights are added into the picture: cosine (metric) similarity is consistently lower than Jaccard similarity  and it is also more volatile for all layers. Table \ref{tab: jac_vs_cos} reports figures for the two most diverse layers. 
\begin{table}[H]
\begin{center}

\subfloat[Unsecured long-term, $J$]{
\begin{tabular}{lllll}
\hline
& 2008& 2009& 2010& 2011 \\
2009 & 61\%*&&&
\\
2010 & 35\%*&42\%*&&
\\
2011 & 18\%*&21\%*&42\%*&
\\
2012 & 15\%*&17\%*&32\%*&70\%*
\\
\hline
\end{tabular}
}
\subfloat[Unsecured overnight, $J$]{
\begin{tabular}{lllll}
\hline
& 2008& 2009& 2010& 2011 \\
2009 & 67\%*&&&
\\
2010 & 53\%*&61\%*&&
\\
2011 & 50\%*&56\%*&71\%*&
\\
2012 & 44\%*&48\%*&60\%*&69\%*
\\
\hline
\end{tabular}
}\\

\subfloat[Unsecured long-term, cosine similarity]{
\begin{tabular}{lllll}
\hline
& 2008& 2009& 2010& 2011 \\
2009 & 29\%*&&&
\\
2010 & 15\%*&16\%*&&
\\
2011 & 2\%*&7\%*&78\%*&
\\
2012 & 1\%*&3\%*&62\%*&89\%*
\\
\hline
\end{tabular}
}
\subfloat[Unsecured overnight, cosine similarity]{
\begin{tabular}{lllll}
\hline
& 2008& 2009& 2010& 2011 \\
2009 & 30\%*&&&
\\
2010 & 13\%*&39\%*&&
\\
2011 & 15\%*&47\%*&52\%*&
\\
2012 & 19\%*&41\%*&48\%*&76\%*
\\
\hline
\end{tabular}
}
\end{center}
\caption{\small{Jaccard similarity and cosine similarity of the same layer of the IIN in different years. The tables refer to the union case. The asterisk denotes statistical significance at 1\% confidence level.}}\label{tab: jac_vs_cos}
\end{table}


A trend towards a greater stabilization seems to arise, since in the lower panels the values along the diagonals are increasing with time. In particular, the cosine similarity of the unsecured long-term layer between 2010-2011 and 2011-2012 is exceptionally high. This corroborates the evidence of a shift on longer maturity, as highlighted in Table \ref{tab:volumes}. The Jaccard similarity between different layers at a certain point in time is relatively low. Considering the intersection, $J$ is around 15-20\% and never exceeds 50\%. This evidence suggests significant complementarity between different segments of the interbank market. The similarity between the overnight unsecured and other layers is at most 30\%, often around 15\%, confirming  that the overnight market is not quite representative of the other layers. Similar results also hold for cosine similarity. \bigskip

 Table \ref{tab: jacc_layers} presents the results of the topological similarity of different layers in 2008 and 2012. Furthermore, the former becomes higher if we restrict to common nodes, meaning that a relatively large number of nodes is present only in one of the two layers. Instead, the similarity between the unsecured layers appears to be less affected by the switch from union to intersection. In general, these results confirm that the network structure differs significantly across different types of contracts.

\begin{table}[htbp]
\begin{center}
\subfloat[2008. Intersection, Union in parenthesis]{
\begin{tabular}{lcccc}
\hline
& S LT & S ST & U OVN & U LT\\
S ST & 18\% (3\%) &&&
\\
U OVN & 12\% (0\%) &15\%* (3\%*)&&
\\
U LT & 5\% (0\%) &13\%* (5\%*) &12\%* (6\%*)&
\\
U ST & 13\% (0\%) &16\%* (4\%*) &29\%* (29\%*) &19\%* (10\%*)
\\
\hline
\end{tabular}
}

\subfloat[2012. Intersection, Union in parenthesis]{
\begin{tabular}{lcccc}
\hline
& S LT & S ST & U OVN & U LT\\
S ST & 26\%* (15\%*)&&&
\\
U OVN & 11\%* (0\%*)& 11\%* (1\%*)&&
\\
U LT & 0\% (0\%) &9\%* (0\%*)& 22\%* (17\%*)&
\\
U ST & 13\%* (0\%*) &11\%* (1\%*)& 32\%* (31\%*) &31\%* (28\%*)
\\
\hline
\end{tabular}
}
\end{center}
\caption{\small{Jaccard similarity between different layers of the IIN in the same year. The asterisk denote statistical significance at 1\% confidence level.}}\label{tab: jacc_layers}
\end{table}

%

\section[null_models]{Results from null models}\label{sec: null_models}

In this section we assess the ability of economically relevant null models to explain the higher order topological properties of the IIN such as reciprocity, clustering and assortativity. While these variables can provide relevant insights in interpreting  the interaction between the network nodes, their value might depend more on the properties of the individual nodes rather than on the properties of the overall network . For example, let us consider the reciprocity concept: in Section \ref{results} we found a reciprocity value of $0.45$ for the total network in 2012. Which value would we find if we allowed each bank to retain the same number of lenders and borrowers as it has in reality, but to randomly choose its counterparties? More generally, which patterns in a real network are ``unexpected'' when one assumes that certain network properties are preserved?

In order to answer these questions, we need to build suitable set of {\it null models} that are maximally random but retain certain network metrics (in the above example, the in- and out- degree of each node). The Maximum Entropy Principle allows to define an \textit{ensemble} of network realizations $G$ and to give a probability $P(G)$ to each of the resulting networks. Specifically, one finds the probability distribution of networks that maximizes entropy, by constraining the average value of some network metrics $\{x_i(G)\}$.  \cite{park2004statistical} show how to build the maximum entropy probability distribution $P(G)$ and to solve for several set of constraints, both for weighted and binary networks. Different constraints lead to different ensembles (and null models).
Appendix \ref{appendixc} provides a detailed explanation of the procedures we followed in building network ensembles, taking advantage of \cite{1367-2630-13-8-083001}, and applications to global trade \citep{2011arXiv1112.2895F} and to credit networks \citep{2013arXiv1302.2063S}.\bigskip

It is useful to introduce the idea of a hierarchy of observables in a network. First order properties of a network involve only linear combinations of the elements of the adjacency or weight matrix. These properties include connectivity and the degree distribution. Analogously one can define second, third, etc. order properties (generically higher order properties) as those metrics that involves sums of products of two, three, etc. elements of the adjacency or weight matrix. The approach is thus to define the ensemble by constraining some low order properties and to observe whether other low order properties and high order properties of the real network are reproduced by the ensemble.\bigskip

In this section we apply the maximum entropy principle to build three ensembles, namely the Directed Binary Configuration model (DBCM), where the in- and out-degree of each node is preserved, the Reciprocal Configuration Model (RCM), where also the number of reciprocated relations of each node is preserved, and the  Directed Weighted Configuration Model (DWCM), where we preserve the in- and out-strenght, as well as the in- and out-degree,  of each node. Among the properties we will test, we consider the number of reciprocated links $R$ (see Eq. \ref{eq: reciprocated}, not for the RCM), the assortativity, the number of triangles $T$ (see Eq. \ref{eq: triangles}).  We will also consider some higher order quantities such as the size of the weakly or strongly connected component and the number of distinct triads or third order motifs, which are the $13$ possible arrangements of up to six directed links among triples of nodes (see top panel of Figure \ref{fig: DBCM_triads}).



\subsection{Directed Binary Configuration Model}

The simplest ensemble we consider is obtained by a model of binary directed networks in which the constrained observables are given by the in- and out-degree sequence of nodes. Following the network literature, we label the ensemble as the \textit{directed binary configuration model} (DBCM).  


\begin{table}
\centering
\subfloat[Unsecured overnight]
{\begin{tabular}{lrrrrr}
\hline
&2008&2009&2010&2011&2012\\
\hline
Largest weak component & 573&565&556&551&532\\
(p-values) & (0.000)&(0.000)&(0.000)&(0.000)&(0.000)\\
Simulation average & 556&547&544&538&515\\
\hline
Largest strong component & 498&486&511&501&456\\
(p-values) & (0.000)&(0.000)&(0.000)&(0.000)&(0.000)\\
Simulation average & 374&359&385&384&335\\
\hline
Reciprocal links & 1,265 &1,231 &1,271 &1,189 &1,033 \\
(p-values) & (0.000)&(0.000)&(0.000)&(0.000)&(0.000)\\
Simulation average & 855&814&843&820&677\\
\hline
Und. triangles & 14,114&11,747&11,645&10,704&10,098\\
(p-values) & (0.000)&(0.000)&(0.000)&(0.000)&(0.000)\\
Simulation average & 18,418&16,252&15,953&14,871&13,755\\
\hline
\end{tabular}
} \\

\subfloat[Secured short-term]
{\begin{tabular}{lrrrrr}
\hline
&2008&2009&2010&2011&2012\\
\hline
Largest weak component &48&48&135&67&35\\
(p-values) & (0.098)&(0.018)&(0.000)&(0.141)&(0.039)\\
Simulation average &52&42&118&72&32\\
\hline
Largest strong component & 14&11&10&17&16\\
(p-values) & (0.033)&(0.030)&(0.063)&(0.000)&(0.000)\\
Simulation average & 11&8&7&9&9\\
\hline
Reciprocal links & 44&44&37&55&42\\
(p-values) & (0.000)&(0.000)&(0.000)&(0.000)&(0.000)\\
Simulation average & 22&14&11&11&11\\
\hline
Und. triangles & 222&132&114&111&72\\
(p-values) & (0.313)&(0.481)&(0.001)&(0.047)&(0.274)\\
Simulation average & 250&137&217&185&91\\
\hline
\end{tabular}
}
\caption{High order properties of two layers of the IIN and the corresponding p-values and average values obtained from simulations of the DBCM.} \label{tab: null1}
\end{table}

For each layer\footnote{With the exclusion of the secured long-term layer which is too small to provide interesting results.} and each year, we solve the extension to directed networks of the system of Eq. (\ref{eq: fermi_system}) in order to obtain the parameters of $P(G)$. Then we simulate a large sample of artificial binary networks and compute the higher order topological properties for each realization of the sample. In this way, we are able to compute both the sample average and the p-values of these properties  that we compare with the values obtained in the real network.  As high order properties we consider the size of the largest weakly and strongly connected component, the number $R$ of reciprocal links, the number $T$ of undirected triangles, the assortativity, and the number of triadic structures.

The first results for two layers are shown in Table \ref{tab: null1}. Looking in the first place at the overnight layer, we see that the selected high order properties of the real network are highly unlikely for a member of the DBCM ensemble. In particular, the size of the largest weak and strong components are much larger than those expected under the null model. The real overnight layer has also a high number of reciprocated links $R$, while the number of undirected triangles $T$ is lower than in the DBCM ensemble. These results are observed also in the unsecured short-term and long-term layers (data not displayed), and they are very stable over time. 
The results for the small secured short-term layer appear instead to be noisier and less stable, with reciprocal links being the only topological property which is always significantly higher than in the DBCM.

A recent investigation of the Dutch interbank market in the period 1998-2008 \citep{2013arXiv1302.2063S} showed that a significant decline of $R$ with respect to the DBCM has anticipated the outbreak of the crisis. They also found that the the number of reciprocated links is significantly smaller than the number predicted by the model. On the contrary, our analysis, based on the period 2008-2012, indicates that during the crisis the reciprocity of all the layers of the IIN has remained quite stable and significantly above the value expected under the DBCM. Clearly, with our dataset we cannot test whether before the crisis the reciprocity was even higher.

Table \ref{tab: assortativity} shows the results of the comparison of assortativity of the Italian interbank layers with that of the DBCM. We see that the disassortative behavior is present also in the DBCM ensemble and that the sample averages are close to the real values. This result is consistent with \cite{RePEc:kie:kieliw:1830} who claim that the assortativity of a network depends on its degree distribution. However our approach shows that in many cases the real layers are significantly more disassortative than expected under a null model preserving the degree distribution. Thus it would be interesting to investigate the origin of these small but significant deviations of disassortativity from null models.

\begin{table}[t]
\centering
\subfloat[Unsecured overnight]
{\begin{tabular}{lrrrrr}
\hline
&2008&2009&2010&2011&2012\\
\hline
Out-degree assortativity & -0.2631&-0.2847&-0.2830&-0.2698&-0.2736\\
(p-values) & (0.001)&(0.001)&(0.004)&(0.013)&(0.004)\\
Simulation average & -0.2467&-0.2668&-0.2682&-0.2593&-0.2566\\
\hline
In-degree assortativity & -0.3466&-0.3563&-0.3748&-0.391&-0.4201\\
(p-values) & (0.003)&(0.005)&(0.015)&(0.019)&(0.003)\\
Simulation average & -0.3273&-0.3366&-0.3587&-0.3738&-0.396\\
\hline
\end{tabular}
} \\

\subfloat[Unsecured long-term]
{\begin{tabular}{lrrrrr}
\hline
&2008&2009&2010&2011&2012\\
\hline
Out-degree assortativity & -0.5263&-0.5395&-0.4549&-0.5208&-0.5141\\
(p-values) & (0.000)&(0.000)&(0.030)&(0.172)&(0.236)\\
Simulation average & -0.4330&-0.4394&-0.4199&-0.5007&-0.5013\\
\hline
In-degree assortativity & -0.3511&-0.3929&-0.236&-0.4391&-0.3803\\
(p-values) & (0.059)&(0.021)&(0.003)&(0.023)&(0.161)\\
Simulation average & -0.3165&-0.3437&-0.1942&-0.4068&-0.3659\\
\hline
\end{tabular}
}
\caption{Assortativity of two layers of the IIN and the corresponding p-values and average values obtained from simulations of the DBCM.}
\label{tab: assortativity}
\end{table}

Finally, we consider the frequency of triads. In \cite{2013arXiv1302.2063S}, authors proposed the use of relative frequency of triads with respect to the expectation of null models as early warning signals of topological collapse of the interbank market. Here we investigate the frequency of triads in different layers of the IIN and we compare them with the DBCM. In the next subsection we will consider the reciprocal configuration model (RCM) as a null model.

In order to use the same procedure for both null models, we employ the software \texttt{MFINDER} which allows to detect network motifs of all orders and to evaluate their frequency against the DBCM or the RCM\footnote{Available at http://www.weizmann.ac.il/mcb/UriAlon/. See also \cite{2003cond.mat.12028M,Milo2002}. The random networks are obtained by rewiring the original links while keeping the number
of incoming edges, outgoing edges and mutual edges of each node. This approach leads to an exact consistency with the constraints, as opposed to the average consistency described in appendix \ref{appendixc}. Therefore we are building microcanonical ensembles rather than (grand) canonical ensembles, as for the other properties. Although the two approaches are exactly equivalent only in the limit of very large networks, the respective linking probabilities converge very rapidly as the number of switches increase. For a detailed comparison see \cite{1367-2630-13-8-083001}.}. In particular, for each triad we compute the z-score, which is equal to the number of observed triads minus the expected value observed from simulations and the result is divided by the standard deviation. Large absolute values of the z-score indicate that the triad is unlikely explained by the null model.

\begin{figure}[htpb]
\begin{center}
\setlength{\tabcolsep}{0pt}
\begin{tabular}{cccccccccccccc}
\includegraphics[width=0.07\textwidth,keepaspectratio=true]{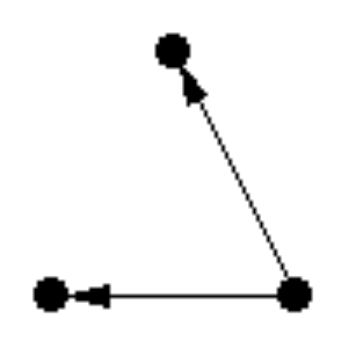} &
\includegraphics[width=0.07\textwidth,keepaspectratio=true]{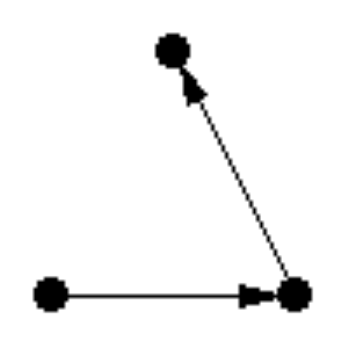} &
\includegraphics[width=0.07\textwidth,keepaspectratio=true]{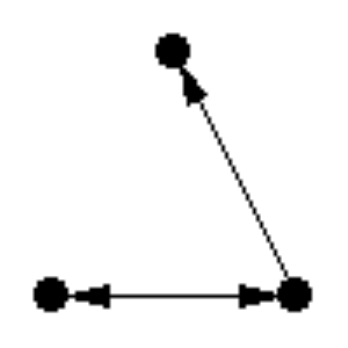} &
\includegraphics[width=0.07\textwidth,keepaspectratio=true]{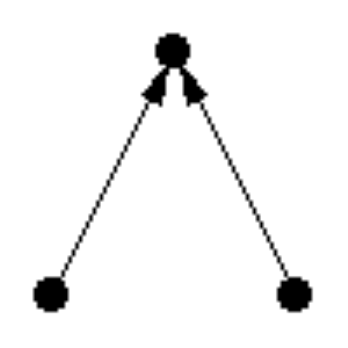} &
\includegraphics[width=0.07\textwidth,keepaspectratio=true]{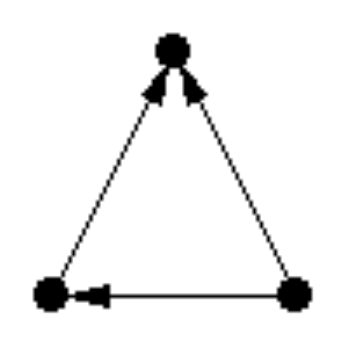} &
\includegraphics[width=0.07\textwidth,keepaspectratio=true]{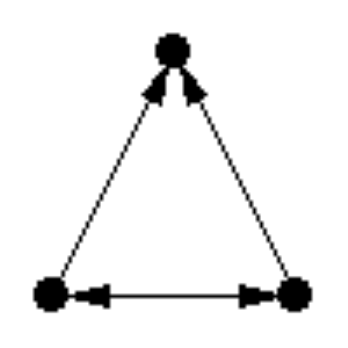} &
\includegraphics[width=0.07\textwidth,keepaspectratio=true]{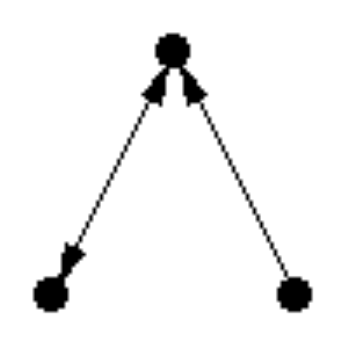} &
\includegraphics[width=0.07\textwidth,keepaspectratio=true]{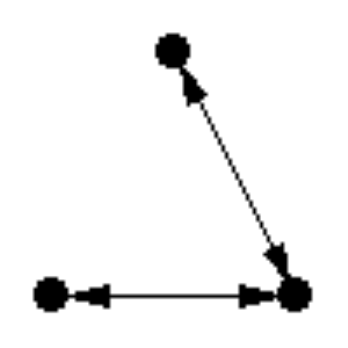} &
\includegraphics[width=0.07\textwidth,keepaspectratio=true]{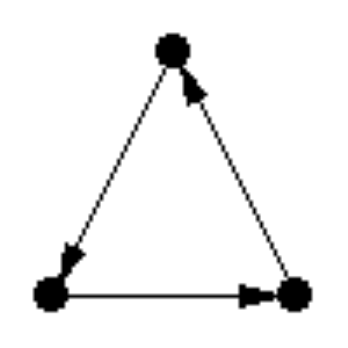} &
\includegraphics[width=0.07\textwidth,keepaspectratio=true]{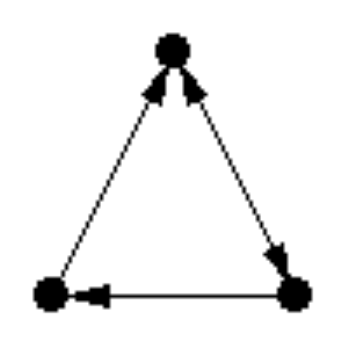} &
\includegraphics[width=0.07\textwidth,keepaspectratio=true]{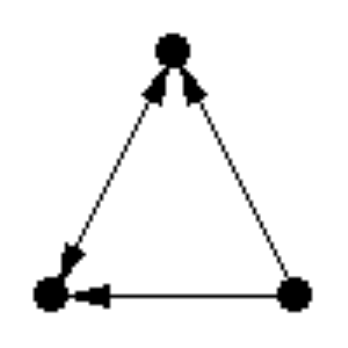} &
\includegraphics[width=0.07\textwidth,keepaspectratio=true]{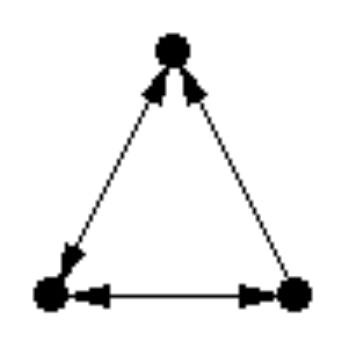} &
\includegraphics[width=0.07\textwidth,keepaspectratio=true]{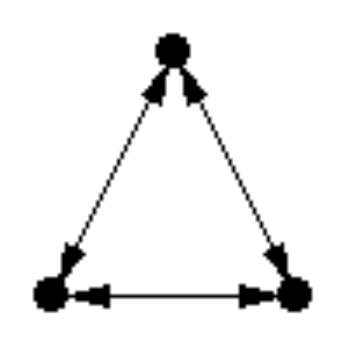} \\
1 & 2 & 3 & 4 & 5 & 6 & 7& 8 & 9 & 10 & 11 & 12 & 13
\end{tabular}\\
\subfloat[Unsecured overnight]
{\includegraphics[width=0.45\textwidth,keepaspectratio=true]{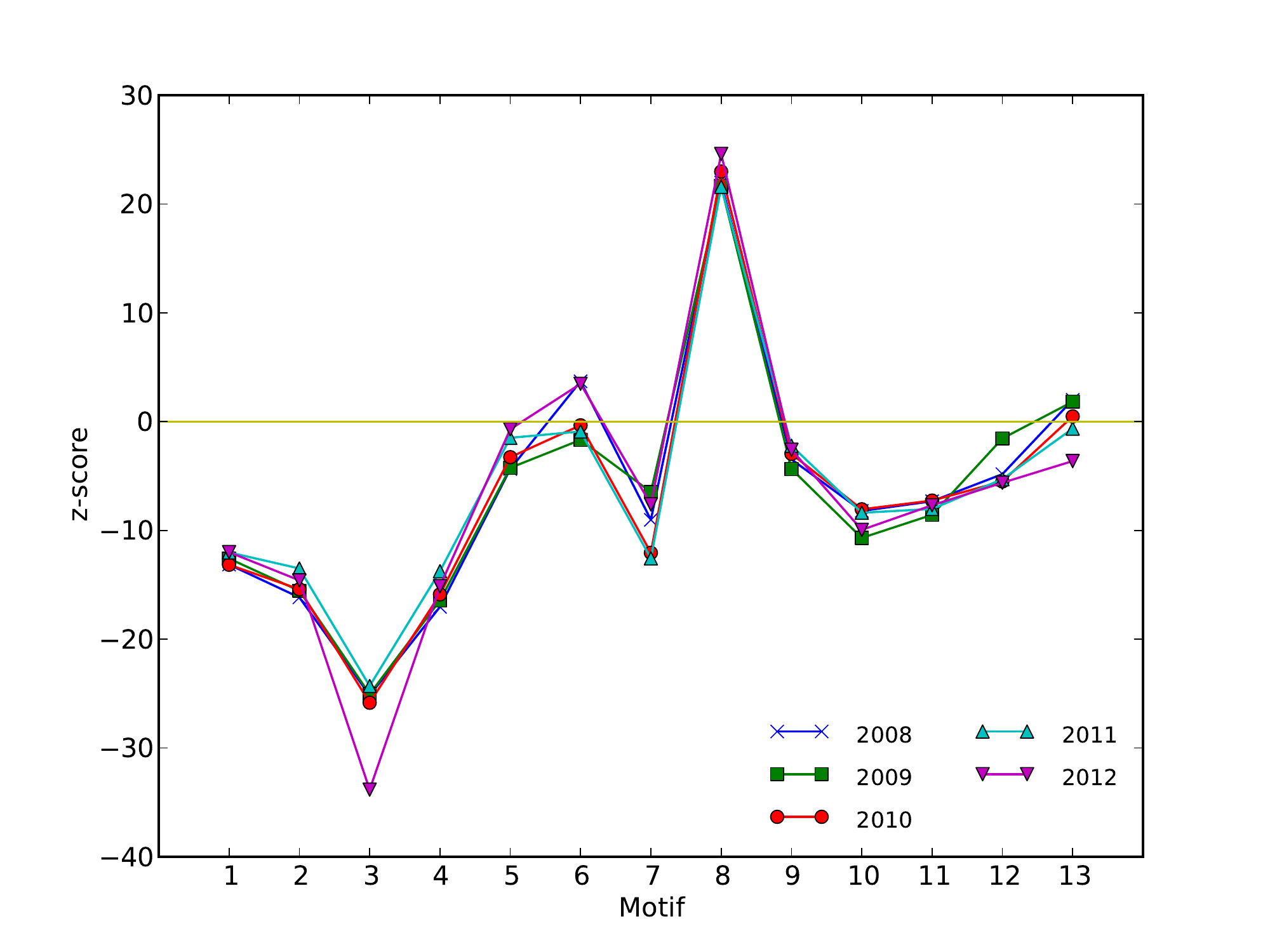}}
\subfloat[Unsecured long-term]
{\includegraphics[width=0.45\textwidth,keepaspectratio=true]{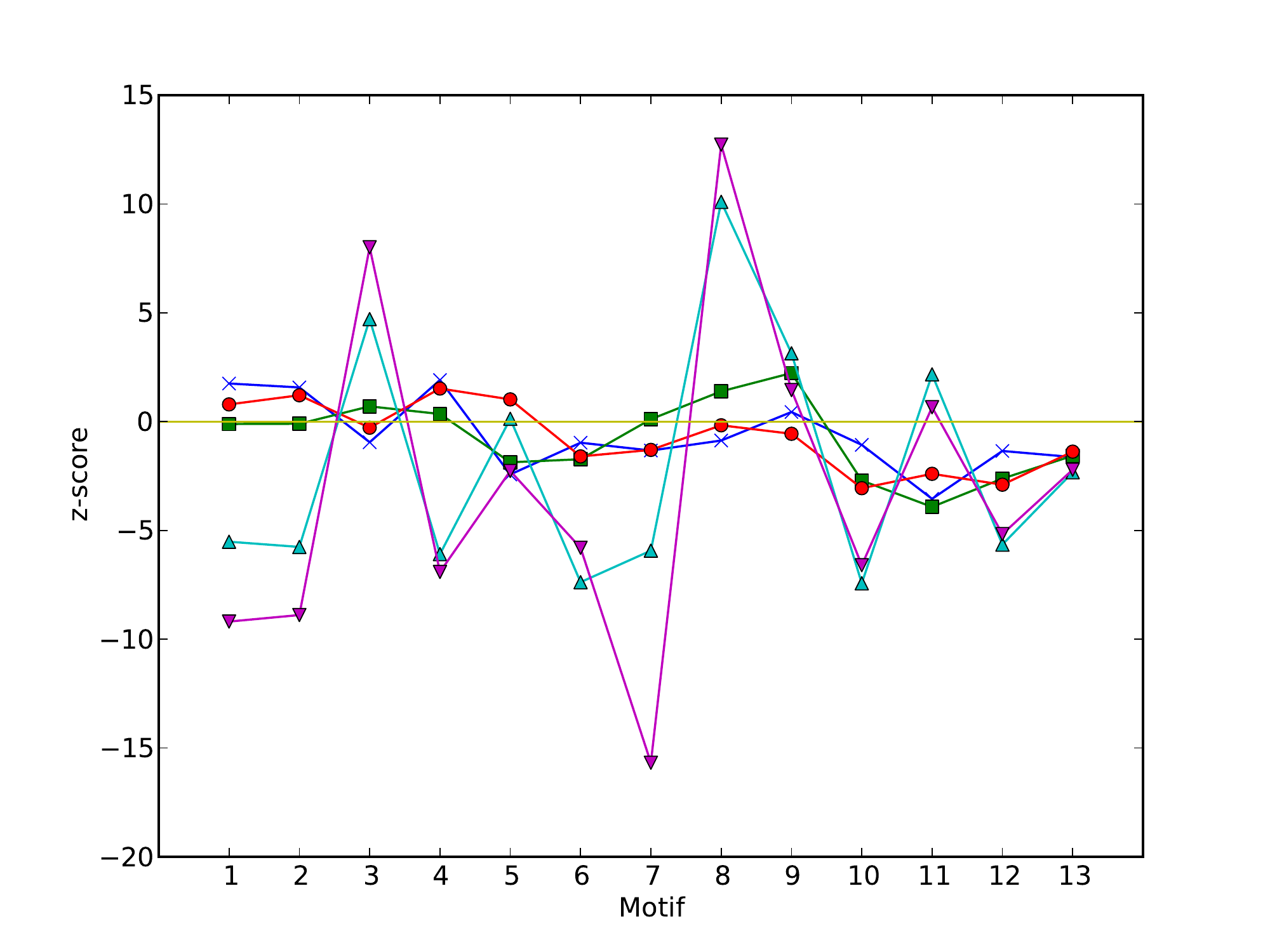}}
\end{center}
\caption{Top panel. The $13$ triads. Bottom panels. Z-scores of the different triads in two layers of the IIN with respect to the DBCM.}
\label{fig: DBCM_triads}
\end{figure}

From Figure \ref{fig: DBCM_triads}, which evaluates triads against the DBCM with the help of z-scores, we see that different layers have different properties. In the overnight layer most triadic structures are under-expressed, and in particular most of those that contribute to the value of $T$ (triads 5 and 9-12). We observe that triad 8, which is the only one strongly over-expressed, is related to the high value of $R$ while at the same time does not contribute to $T$. Similar properties are exhibited by the unsecured short-term layer (left panel of Figure \ref{fig: U*ST_triads}) and by the total network (data not displayed), which again reflects closely the properties of the overnight layer. The z-scores of the unsecured long-term layer are instead unstable across years, with most triads not clearly under- or over-expressed across the period, although we still find a tendency to under-express triads 6 and 10-13, consistently with the low value of $T$ \footnote{The small secured short-term layer exhibits 
similar properties.}. In conclusion, this analysis highlights that the pattern of over- or under-abundance of triads with respect to the DBCM is different in different layers of the IIN. Also the time stability of these patterns appears to be quite different in different layers. 

\subsection{Reciprocal Configuration Model}



As we have discussed above, reciprocity plays an important role that it is not captured by the DBCM. This fact can have a significant effect on the analysis of triads described above.  In order to control for the effect of $R$ on the frequency of triads, \cite{2013arXiv1302.2063S} introduce a more complex null model, i.e. the reciprocal configuration model (RCM). In this ensemble $R$ takes on average the same value of the real networks under study.

\begin{figure}[htpb]
\begin{center}
\subfloat[Unsecured overnight]
{\includegraphics[width=0.45\textwidth,keepaspectratio=true]{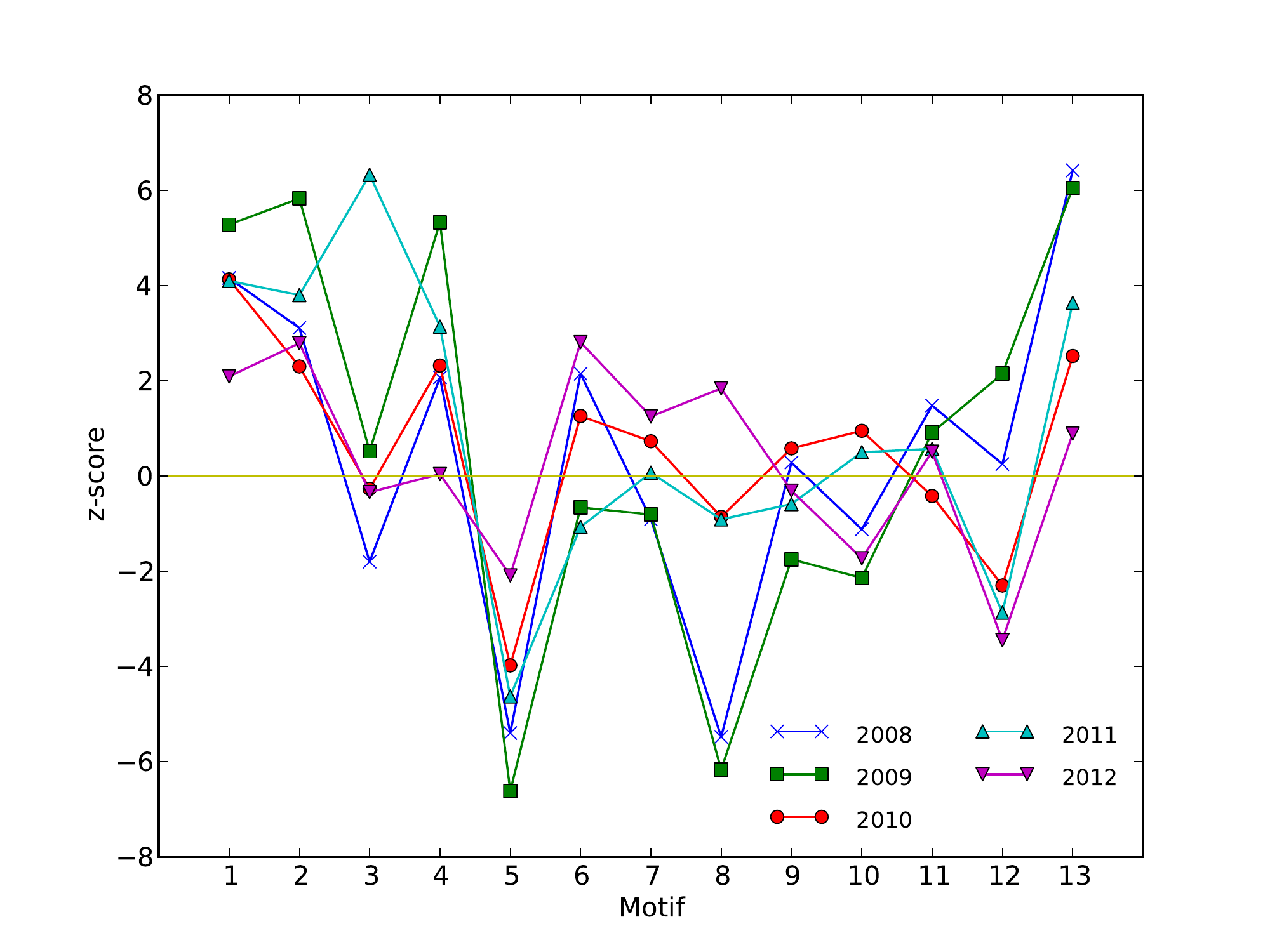}}
\subfloat[Unsecured long-term]
{\includegraphics[width=0.45\textwidth,keepaspectratio=true]{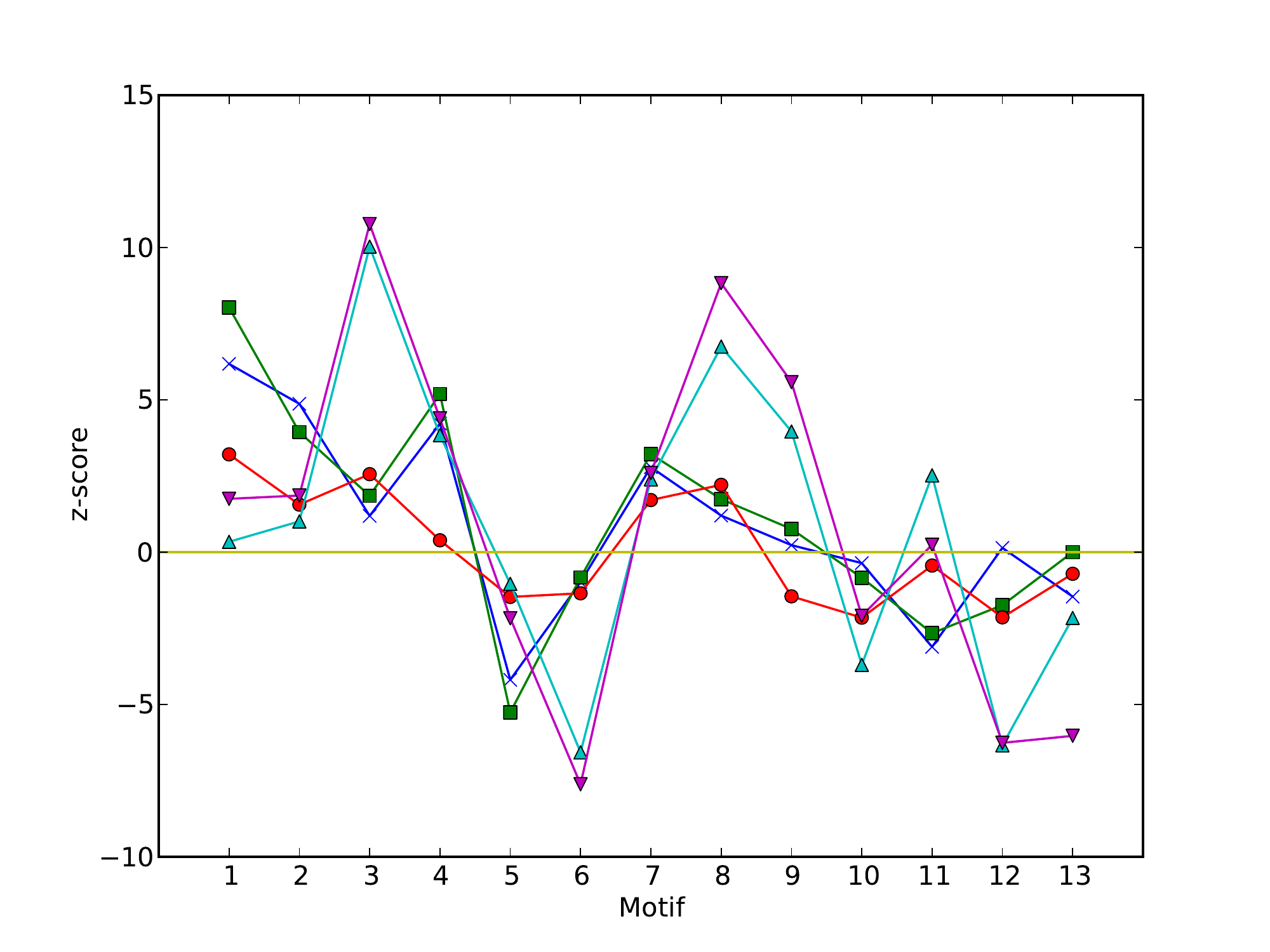}}
 \end{center}
\caption{Z-scores of the different triads in two layers of the IIN with respect to the RCM. See the top panel of Fig. \ref{fig: DBCM_triads} for the correspondence between the x-axis and the triads.}
\label{fig: RCM_triads}
\end{figure}

\begin{figure}[htpb]
\begin{center}
\subfloat[DBCM]
{\includegraphics[width=0.45\textwidth,keepaspectratio=true]{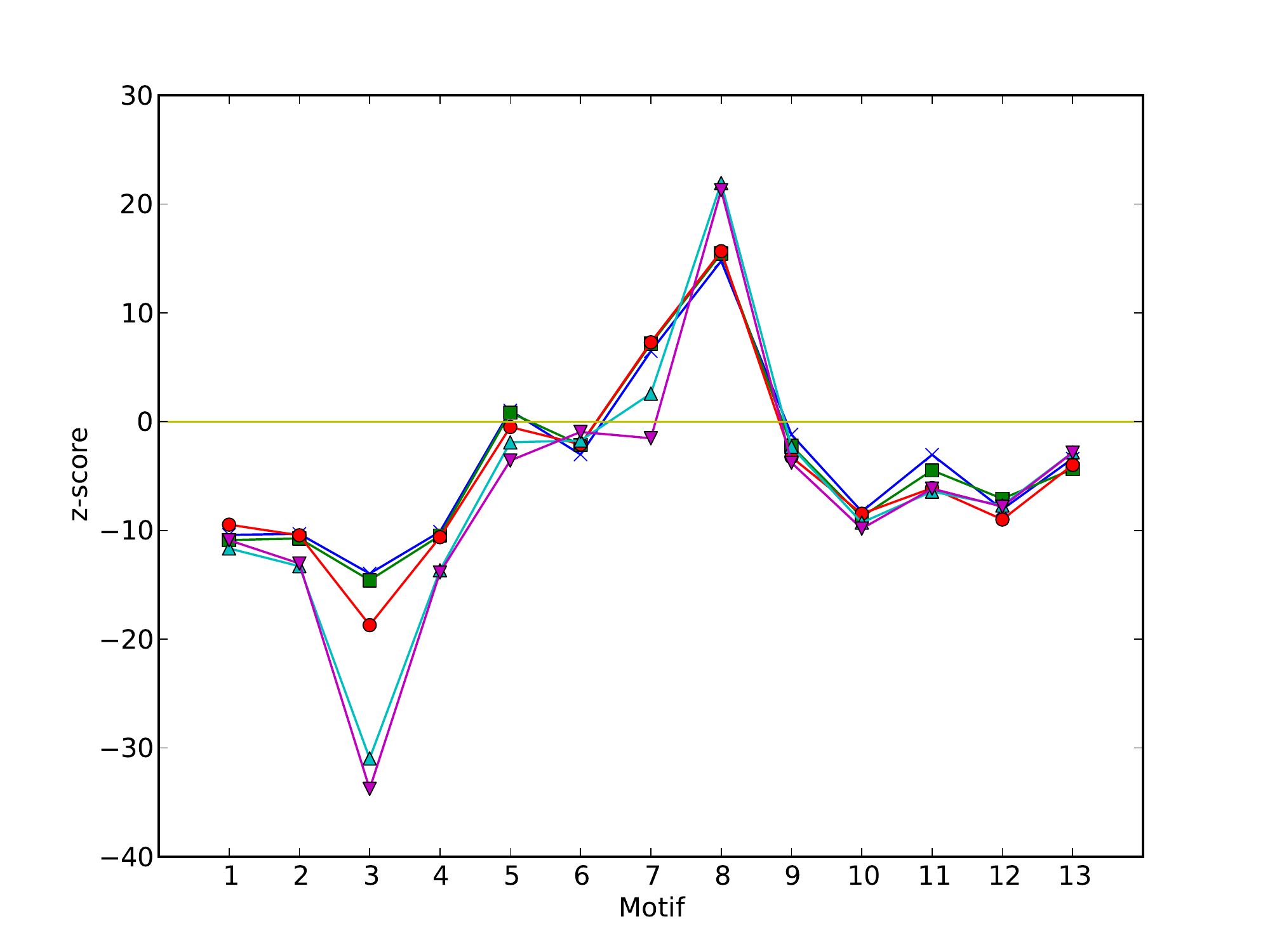}}
\subfloat[RCM]
{\includegraphics[width=0.45\textwidth,keepaspectratio=true]{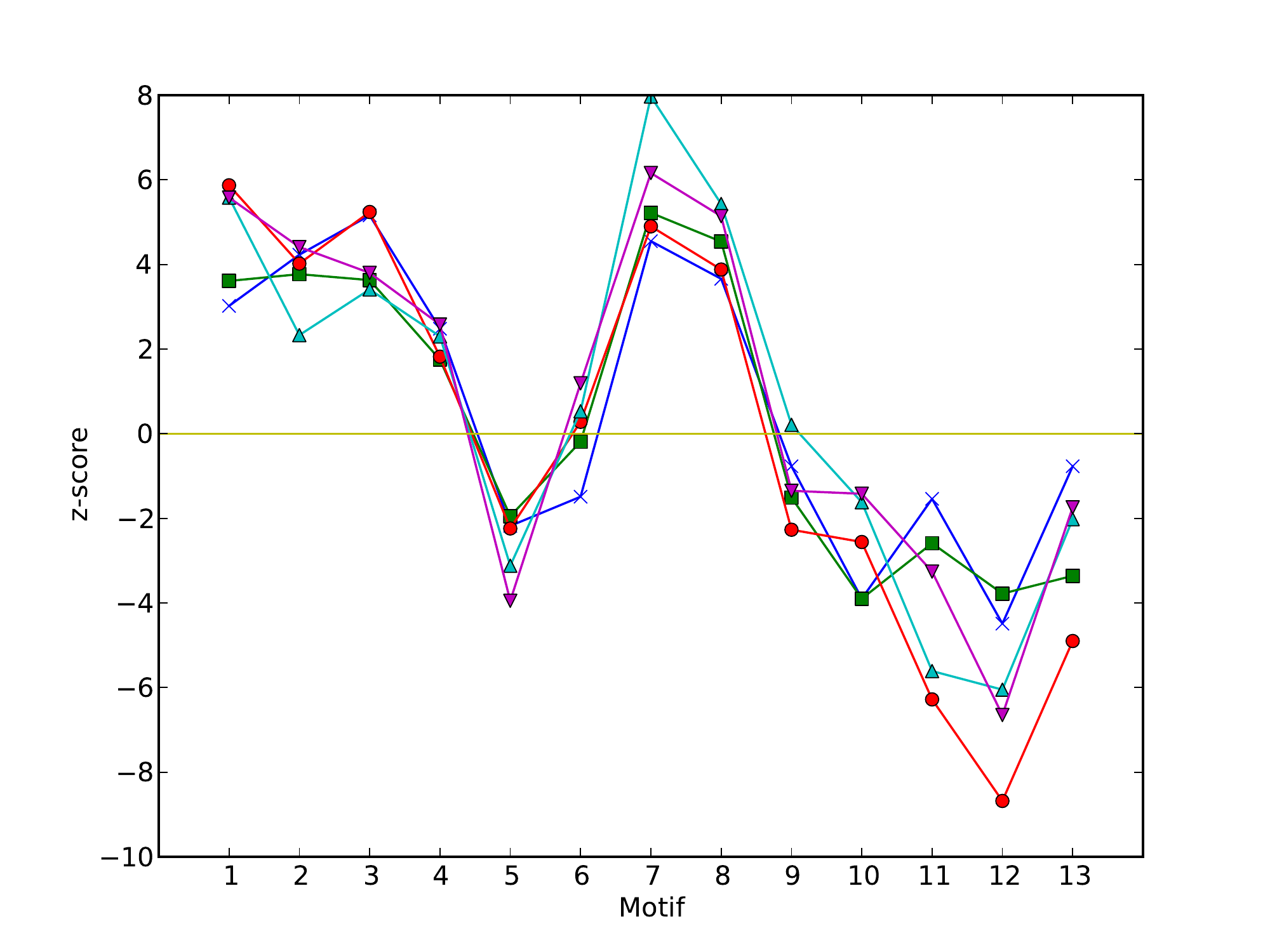}}\\
 \end{center}
\caption{Z-scores of the different triads in the unsecured short-term layer with respect to the DBCM (left panel) and RCM (right panel). See the top panel of Fig. \ref{fig: DBCM_triads} for the correspondence between the x-axis and the triads.}
\label{fig: U*ST_triads}
\end{figure}

In Figure \ref{fig: RCM_triads}, we present the z-scores of triads, for the same layers, evaluated against the RCM  and obtained again by using again the software \texttt{MFINDER}. Firstly, we observe that the z-scores decrease drastically for the unsecured overnight layer with respect to the DBCM. As a consequence, the under- or over-expression of triads turns out to be significant only in some years. We regard this result as a consequence of the dependency of higher order properties on lower order properties for this layer. The decline of z-scores occurs also for the unsecured short-term layer (Figure \ref{fig: U*ST_triads}), although in this case we still observe a much stabler and pronounced deviations of triads from the RCM. Additionally, in this layer the z-scores of triads are affected in a very different way when we pass from the DBCM to the RCM: triads 1-4 become over-expressed while in the DBCM they are under-expressed; triads 5 and 7 get more distant from the null model; triads 6, 8-13 are fairly 
similar to the DBCM. Regarding the unsecured long-term layer, we
observe instead no systematic decrease of the z-scores. In particular, the values for triads 3, 5-6, 8-12 are fairly unaffected, while triads 1,2,4 become under-expressed as opposed to over-expressed. Only for the dyad 7 we observe some significantly lower z-scores. 

From these results we learn that the influence of dyads on triads is anything but linear. We conclude that, while the RCM provides a better statistical representation of the topology of some real interbank layers than the DBCM, it does not explain completely their third order properties. Moreover, as for the DBCM, the pattern of over- and under-expression of triads as well as their time stability is quite different in different layers of the IIN.

\subsection{Directed Weighted Configuration Model}

In the remaining of this section we focus on a null model for weighted and directed networks with given strength and degree distributions, which we call the \textit{directed weighted configuration model} (DWCM). 
The technical details regarding this model are presented in the Appendix \ref{appendixc}. The main idea can be summarized as follows. The standard ensembles for weighted networks do not have a realistic topology, since the only constraints entering these models are the values of the strength distribution. In particular, they have a much higher connectivity than their real counterparts. Furthermore, as we have seen from Table \ref{tab: spearman}, in our layers the degree distribution is highly correlated with the strength distribution. 
Note that this correlation can appear only in sparse networks, since in the limit of a dense topology all nodes converge to the same degree value, regardless of their strength. By resorting to a model where strengths and degree act jointly as constraints, we take into account both the strength-degree correlation and the sparsity of the original network \footnote{From an economic viewpoint, it is useful to remark that strengths are not by themselves network variables, rather they reflect  economic variables, such as the economic size of nodes or their preferences. Thus we can argue in favor of an indirect connection between economic behavior and topology, which to might help to explain why, in a world of heterogeneous agents facing unpredictable events, it is generally impossible to determine a universal statistical law governing the degree distributions of different economic networks, as well as of the same network over time (Section \ref{results}). This lack of universality supports the modeling approach
followed in this section, which does not require to make
hypotheses of
this sort.}.

We generate the ensemble of networks and we consider the metric version of some of the higher order properties considered above (see appendix \ref{appendixa} for definitions). From Table \ref{tab: DWCM} we see that the strength reciprocity is often explained by the null model, although there is some propensity of the overnight layer to display larger values than the ensemble average while the opposite holds for the long-term layer. To complete the picture, we find that the values of the other layers (not displayed) are substantially in line with the null model. These results indicate that in the IIN the net exposure $ \Delta_{ij} = w_{ij} - w_{ji}$ between couples of banks is mostly determined by their respective out- and in-strengths. Instead, there is a clear orientation of layers (including those not displayed) to be less disassortative than the null model, which may be potentially of interest in terms of systemic stability. In fact, assortativity in this case means that banks are more likely to have
large credit (debt) positions with banks which also
have large credit (debt) positions, potentially compounding the risk of contagion in case of default. In this sense, our results can be interpreted as an indication that for a set of alternative ``realistic'' network configurations the contagion risk could be reduced.

\begin{table}
\centering
\subfloat[Unsecured overnight]
{\begin{tabular}{lrrrrr}
\hline
&2008&2009&2010&2011&2012\\
\hline
Out-strength assortativity & -0.0204&-0.0221&-0.0334&-0.0367&-0.0316\\
(p-values) & (0.000)&(0.000)&(0.000)&(0.000)&(0.000)\\
Simulation average & -0.1458&-0.269&-0.3016&-0.3328&-0.2711\\
\hline
In-strength assortativity & -0.0378&-0.0528&-0.1551&-0.197&-0.1827\\
(p-values) & (0.000)&(0.000)&(0.000)&(0.000)&(0.000)\\
Simulation average & -0.2739&-0.2527&-0.2817&-0.2623&-0.2543\\
\hline
Strength reciprocity & 0.4325&0.1407&0.1157&0.0714&0.2070\\
(p-values) & (0.000)&(0.124)&(0.498)&(0.149)&(0.004)\\
Simulation average & 0.1929&0.1078&0.1158&0.0912&0.1334\\
\hline
\end{tabular}
} \\

\subfloat[Unsecured long-term]
{\begin{tabular}{lrrrrr}
\hline
&2008&2009&2010&2011&2012\\
\hline
Out-strength assortativity & -0.0674&-0.0328&-0.058&-0.1231&-0.3209\\
(p-values) & (0.070)&(0.022)&(0.000)&(0.000)&(0.000)\\
Simulation average & -0.2038&-0.1989&-0.2223&-0.4768&-0.4649\\
\hline
In-strength assortativity & -0.0676&-0.0364&-0.0407&-0.0222&-0.0360\\
(p-values) & (0.038)&(0.000)&(0.025)&(0.000)&(0.000)\\
Simulation average & -0.2140&-0.3011&-0.1300&-0.2618&-0.2808\\
\hline
Strength reciprocity & 0.0596&0.0079&0.0004&0.0058&0.0139\\
(p-values) & (0.065)&(0.202)&(0.018)&(0.000)&(0.002)\\
Simulation average & 0.0236&0.0530&0.0192&0.0346&0.0287\\
\hline
\end{tabular}
}
\caption{High order properties of two layers of the IIN and the corresponding p-values and average values obtained from simulations of the DWCM.}
\label{tab: DWCM}
\end{table}

\begin{figure}[htbp]
 \centering
 \includegraphics[width =0.5\textwidth,keepaspectratio = True,angle = 90]{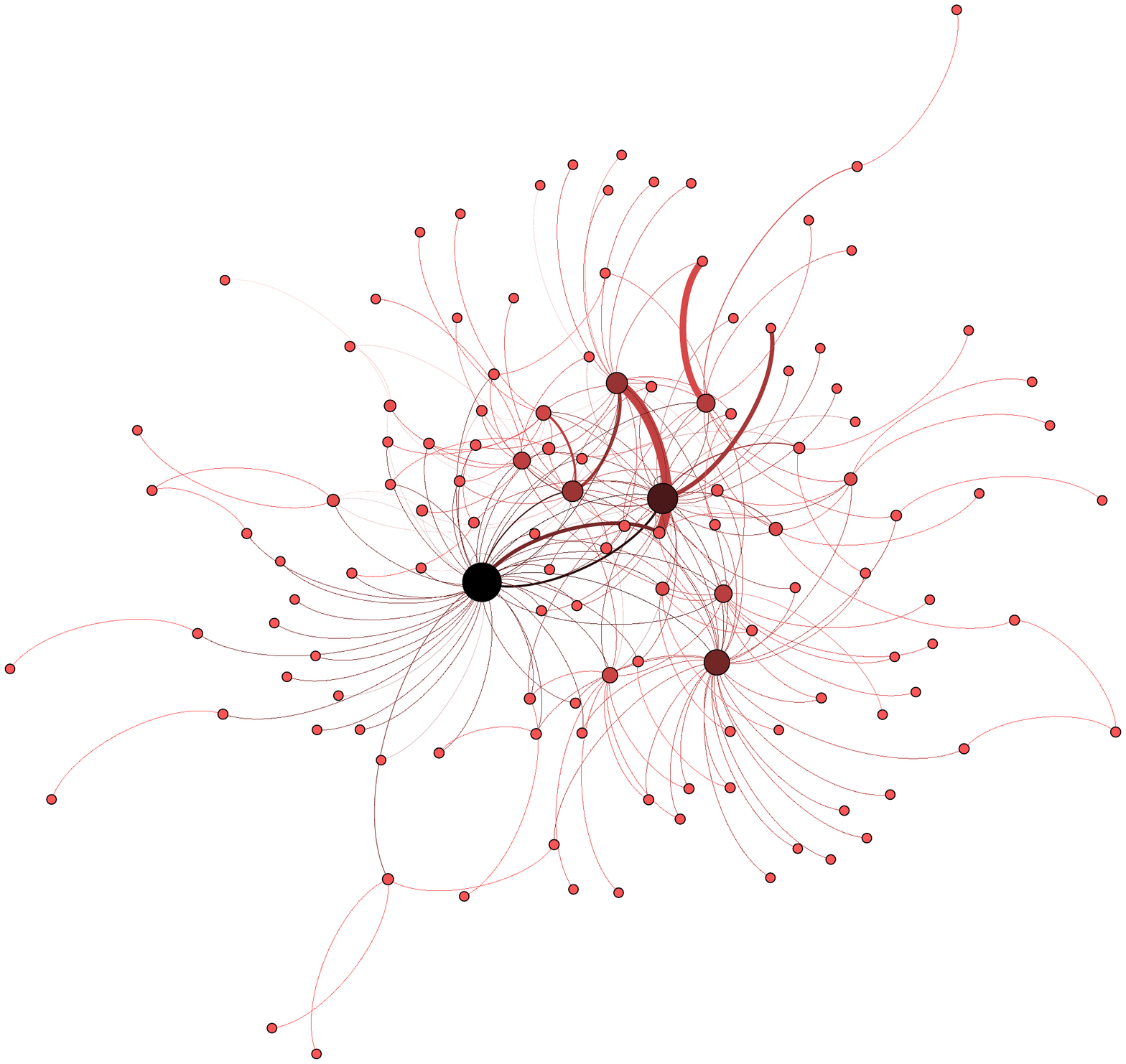}
 \caption{A realization of a random network from the DWCM of the unsecured long-term in 2009. Note that the direction of links is clockwise; the size of nodes is proportional to their strength; the thickness of arcs is proportional to their values; a darker color is associated to nodes with the largest degree.}
\label{fig: DWCM_net}
\end{figure}

Finally, an important insight offered by the DWCM regards a widely investigated property of credit networks, namely their core-periphery structure \citep{RePEc:fip:fedcwp:0912,RePEc:kie:kieliw:1759,RePEc:dnb:dnbwpp:348}. It is possible to show analytically that the subdivision of nodes into a core, made of nodes highly connected with the other core members as well as with peripheral nodes, and a periphery, made of nodes with no reciprocal connections, depends entirely on the degree distribution of the network \citep{2011arXiv1102.5511L}. 
Accordingly, from Figure \ref{fig: DWCM_net} we see that a simulation of a  network from DWCM  can display a clear core-periphery structure\footnote{This is true also for DBCM networks, because the topology of the two ensembles is identical.}. Since in this model, and likewise in real credit networks, degrees are correlated with strengths, which on their part are
economically linked with bank size, we get to the heuristic conclusion that the core-periphery subdivision is related to the concentration of the credit market and not to some additional topological property.

\section{Conclusions}\label{sec:conclusions}

Network theory is increasingly employed for systemic risk assessment, as a network topology and its dynamics are key in determining the likelihood and the severity of contagion episodes. This work accounts for the complexity of modern interbank markets and provides a broad analysis of its different segments (layers) over an interesting time lapse, as the 2008-2012 period featured several crisis episodes. Italian interbank markets responded in several ways: while a significant shift from short term to longer maturities took place, at the same time the domestic overnight money market displayed a significant resilience. \bigskip

Topological properties differ significantly across layers. The topology of the overall network largely reflects the one of the unsecured overnight segment. In this layer, the persistence of credit relationships over time (the probability that two banks have a link at time $t$ provided that they had a link at time $t-1$) is significantly larger with respect to other segments, while similarity across layers is notably low. Such results may provide an input for future work on interconnectedness and contagion, as they might help assessing the severity of data limitation cases. We also put forward a comprehensive analysis of random models for the estimation of linkages in interbank markets: higher order topological properties (the so-called network motifs) of some layers differ from those of a random network, suggesting sophisticated network models might be needed to provide a comprehensive representation of credit markets.  \bigskip

From a policy perspective, the heterogeneity displayed by different layers may be good news for financial stability, as it is likely to slow contagion across institutions. Evidence that the overnight unsecured market  most closely mirrors the topological features of the overall (total) network should provide comfort to policymakers, as the overnight unsecured interbank market is the focus of monetary policy operations in several jurisdictions. Still, if policymakers and researchers alike were to target a specific segment of the interbank network, they should be careful in adopting an analytical framework based on the overall features of the network.

\appendix

\section{Appendix A: Network Statistics Definition\protect\bigskip}\label{appendixa}

In order to represent interbank networks we refer to the graph $G=(V,E)$, where $V$ is the vertex or node set, typically assumed to be a subset of $\field{N}$, while $E$ is the edge or link set, with $E\subset\field{N}\times\field{N}$, where $(i,j) = e_{ij} \in E $ can eventually map onto a subset $D$ of $\field{R}$. In this case $w(e_{ij}) = w_{ij} \in D$ is said to be the strength of the link $(i,j)$ and $G$ is said to be \textit{weighted}. If $(i,j)$ maps onto ${0,1}$ we say that the network is \textit{binary} or \textit{unweighted}. Further we set $n = |V|$ and $l = |E|$. We say that $G$ is undirected or that it is a graph if we suppose $e_{i,j}\equiv e_{j,i}$ for each $i,j\in V$. Otherwise we say that $G$ is directed or that it is a digraph. In general, $G$ may be represented by the adjacency matrix $A$ with elements $a_{ij} = 1$ if $e_{ij} \in E$ and $a_{ij} = 0$ otherwise. If $G$ is undirected, $A$ is symmetric. The strength between nodes may be represented by the matrix $W$ with elements $w_{ij}$.
From our definitions we see that a weighted network can be always \textit{projected} onto a a binary network by setting $a_{ij} = 1$ if $w_{ij} > 0$. Further, directed binary networks may be \textit{symmetrized} using the following update rule: $a_{ij} \leftarrow a_{ij} + a_{ji} - a_{ij}a_{ji}$. 

The \textit{neighborhood} of a node $i$ is defined as the set $\psi(i)$ of nodes such that $e_{ij} \in E$ for all $j \in \psi(i)$ (the definition may be easily adapted for directed networks). Then the \textit{degree} $k_i$ is the cardinality of $\psi(i)$.
%
In digraphs we must distinguish between the \textit{in-degree} $k_{i}^{in}$, i.e. the number of edges pointed to some vertex, and the \textit{out-degree} $k_{i}^{out}$, i.e. the number of edges pointing away from it.
%
%
By extension, the \textit{strength} of a node may be defined as follows:

\begin{equation}
w_i= \sum_{j\in \psi(i)} w_{i,j}
\end{equation}

with obvious extension to the directed case.

In directed networks, such as the interbank networks, it's interesting to measure the likelihood of double links (with opposite directions) between nodes pairs. In the simplest form, the number of reciprocated links $R$ is given by the sum

\begin{equation}\label{eq: reciprocated}
R = \frac{1}{2}\sum_i \sum_{j \neq i} a_{ij} a_{ji} 
\end{equation}

%
%
%
%
The preferred measure of \textit{reciprocity} in the literature \citep{soramaki2007topology} is the correlation coefficient of $A$ and $A^T$:

\begin{equation} \label{eq:recip}
\rho = \frac {\sum_{i \neq j} (a_{ij} - \bar{a}) (a_{ji} - \bar{a})}{\sum_{i \neq j} (a_{ij} - \bar{a})^2} 
\end{equation}

where  $\bar{a}$ is the average value of the entries of $A$. These measure may be adapted easily to strengths:

\begin{equation}\label{eq:w_recip}
\rho_w = \dfrac{\varrho_w - \bar{w}}{\omega - \bar{w}}
\end{equation}

where

\begin{equation} 
\varrho_w = \dfrac{\sum_{i \neq j}w_{ij}w_{ji}}{\sum_{i \neq j}w_{ij}} \qquad \qquad 
\omega =  \dfrac{\sum_{i \neq j}w_{ij}^2}{\sum_{i \neq j}w_{ij}}
\end{equation}

We observe that, since reciprocity is defined only for $i \neq j$, selfloops are dropped by construction when we compute this measure.

A network is said to be assortative if the degree of a node is positively correlated with the degree of its neighbors. Otherwise, it may be disassortative or uncorrelated. The \textit{assortativity coefficient} is essentially the Pearson correlation coefficient of $k$ between pairs of linked nodes.
%
%
%
Alternatively, we can capture assortativity by examining the average degree of neighbors of a node with given degree. This value may be written as $\langle k_{nn} | k \rangle = \sum_{k' \in D'}{k'P(k'|k)}$, where $P(k'|k)$ is the conditional probability that a node with degree $k$ points to a node with degree $k'$. If $\langle k_{nn} |k \rangle$ is increasing in $k$, the network is assortative. 

%
%
%
A second set of measures is related to connectivity. The simplest example is given by \textit{density} which 
%
%
for directed graphs reads

\begin{equation}
d = \frac{l}{n(n-1)}
\end{equation}

Interbank networks are found to display low density, i.e. to be sparse. A network is said to be \textit{sparse} when $\lvert E \rvert \ll \lvert V \rvert^{2} $. Sparsity is a fundamental shared property of real networks. Sparsity cohabits with another very common property, i.e. the fact that most nodes in a network are connected by a path made of consecutive links\footnote{A path in a graph is a sequence of vertices such that from each of its vertices there is an edge to the next vertex in the sequence.Two vertices $i$ and $j$ are said to be connected if $G$ contains a path from $i$ to $j$. A \textit{connected component} is a maximal connected subgraph of $G$.}. In this case we also say that $G$ has a giant component, meaning that most of the nodes lie on a single component. This property is generally detected also in interbank network.

The \textit{distance} $d_{ij}$ between two nodes $i$ and $j$ may be defined as the shortest (geodesic) path between $i$ and $j$ when $i \neq j$. If $i$ and $j$ are not connected, we set $d_{ij} = + \infty$. Then the average distance $\bar{d}$ must be computed separately for each of the connected components of $G$. For directed networks, $d_{ij}$ is computed on the symmetrized network (see above).


In a binary directed network the \textit{undirected} clustering coefficient of a node $i$ is the standard clustering coefficient of its symmetrization:

\begin{equation}  \label{eq: und_cc}
ucc_i=\frac{\sum_{h \neq j} (a_{ij} + a_{ji} - a_{ij}a_{ji}) (a_{jh} + a_{hj} - a_{jh}a_{hj}) (a_{ih} + a_{hi} - a_{ih}a_{hi}) }{k_i (k_i - 1)}
\end{equation}

Instead, the \textit{directed} clustering coefficient of a node $i$ in the same network is defined as follows \citep{PhysRevE.76.026107}

\begin{equation} \label{eq: dir_cc}
dcc_i=\frac{\sum_{h \neq j} (a_{ij} + a_{ji}) (a_{ih} + a_{hi}) (a_{jh} + a_{hj})}{ 2 (k_i (k_i - 1) - 2 k_i^{\leftrightarrow})}
\end{equation}

where $k_i$ is the total degree of $i$, i.e.  $k_i = k_i^{out} + k_i^{in}$ and $k_i^{\leftrightarrow} = \sum_j a_{ij} a_{ji}$ is the number of bilateral links. Self-loops are excluded from both computations. Finally, the number of undirected triangles in a directed network, which is used in sec. \ref{sec: null_models}, may be defined, using the symmetrization rule introduced above, as follows:

\begin{equation} \label{eq: triangles}
T =\frac{1}{3} \sum_i \sum_{j \neq h} \left[ (a_{ij} + a_{ji} - a_{ij}a_{ji}) (a_{jh} + a_{hj} - a_{jh}a_{hj}) (a_{ih} + a_{hi} - a_{ih}a_{hi}) \right]
\end{equation}

\section{Appendix B: Similarity Analysis Methodology\protect\bigskip}\label{appendixb}

Generally speaking, different measures of similarity are convenient for different types of analysis.
For instance, we wish to have differentiated measures for binary and numerical data.
Since graphs can be represented in alternative ways, in network theory a variety of measures have been adopted to measure network similarity, which borrow from different fields.
A common requirement is that the similarity measure $s$ is related to a metric distance $d$ by some simple relation like $s + d =  k$ for some constant $k$.

The most widely used similarity coefficients for valued vectors are the cosine and the Pearson correlation coefficient. 
Cosine similarity is defined as

%
%
\begin{equation} \label{eq: cosine}
\cos (\theta) =  \frac{ \mathbf{p} \cdot \mathbf{q}} { \|p\| \|q\|} = \frac{ \sum\limits_{i=1}^{n}{p_i  q_i} }{ \sqrt{\sum\limits_{i=1}^{n}{p_i^2}} \sqrt{\sum\limits_{i=1}^{n}{q_i^2}}} 
\end{equation}

This coefficient takes values in the interval $[-1,1]$ ($[0,1]$ if the vectors are nonnegative). Here $\theta$ is the angle formed by $\mathbf{q}$ and $\mathbf{p}$. 
%
%
Pearson correlation is identical to the cosine of centered data. 
The main disadvantage of correlation as a network similarity measure is that it assumes that the entries of the data vector are equally distributed, while this might not be the case in a network (see section \ref{sec: null_models}). We might think to account for the heterogeneity of nodes in the network by centering data with respect to the ME expectation (\ref{eq: me_expect}). We prefer instead to make our measure independent from distributional assumptions, and consequently choose (\ref{eq: cosine}) as our similarity measure for weighted data.

Regarding boolean data, the most widely used similarity measures are the Jaccard similarity and Dice or S\o{}rensen similarity. Given two binary vectors, the Jaccard similarity is

\begin{equation} \label{eq: jaccard}
J(\mathbf{p},\mathbf{q})  =  \frac{|\mathbf{p} \wedge \mathbf{q}|}{|\mathbf{p} \vee \mathbf{q}|}
\end{equation}

Here $\wedge$ ($\vee$) stands for the entry-wise maximum (minimum) of $\mathbf{p}$ and $\mathbf{q}$. Dice similarity $D$ is related to $J$ by the relationship $D = 2J/(1+J)$, but the distance derived from $D$, differently from $J$, is not a proper metric.

In graph theory, the Edit distance is also used frequently. This metric has the disadvantage of depending on imposition of rather restrictive conditions over an unidentified cost function. Alternatively, the following similarity coefficient has been proposed recently \citep{Bunke:1998:GDM:289720.289729}:

\begin{equation}
\frac{|G^*|}{\max(|G_1|,|G_2|)}
\end{equation}

where $|G|$ stands for the number of nodes of graph $G$, and $G^*$ is the maximal common subgraph between $G_1$ and $G_2$. The disadvantage of this approach is that finding $G^*$ is computationally expensive \citep{Dehmer2006447}. Furthermore, $|G^*|$ may be largely determined from errors and distortion in data, especially for weighted networks \citep{Bunke:1998:GDM:289720.289729}. Given the mentioned shortcomings of alternative similarity measures, we choose to employ $J$ as argued in section \ref{sec:similarity}.

\section{Appendix C: Null Models Methodology\protect\bigskip}\label{appendixc}

In  this appendix we summarize briefly the methodology of \cite{park2004statistical}. As stated in the main text (sec. \ref{sec: null_models}), the average network observables are defined in terms of statistics computed over the network ensemble. Since the observables depend on network realizations, we weight the average against the probability $P(G)$ of observing a given realization $G$ in the ensemble:

\begin{equation} \label{eq: general_system}
\langle x_i \rangle = \sum_{G \in \mathcal{G}} P(G) x_i(G) = \bar{x_i}
\end{equation}

Since the $x_i(G)$ are a given, we need to specify a parameter dependent functional shape of $P(G)$ in order to solve the system. By adopting the basic concepts of equilibrium statistical mechanics we obtain a solution for this task by maximizing the following Lagrangean:

\begin{equation} \label{eq: lagrang}
\mathcal{L} =  S + \lambda ( 1 - \sum_{G}P(G)) + \sum_i \theta_{i} \left ( \bar{x_i} - \sum_G P(G) x_i(G) \right )
\end{equation}

where $S = - \sum_G P(G) \ln P(G)$ is Gibbs entropy. By taking the f.o.c. we obtain

\begin{equation}
\ln P(G) + 1 + \lambda + \sum_i \theta_i x_i(G) = 0
\end{equation}

Rearranging and taking antilogs:

\begin{equation} \label{eq: boltzgibbs}
P(G) =  \dfrac{e^{-H(G)}}{Z}
\end{equation}

where $H(G) \equiv \sum_i \theta_i x_i(G)$ is the graph Hamiltonian which, thanks to matrix representation of $G$, can be rewritten in terms of the matrix $W$ or $A$, and $Z \equiv e^{(\lambda + 1)}$ is the partition function. From the normalization constraint we easily obtain that $ Z = \sum_G e^{-H(G)}$. The model is solved when the values of the parameters $\{\theta_i\}$, which fully determine $P$, are obtained from the system (\ref{eq: general_system}). It is possible to show that, if we adopt the Boltzmann-Gibbs distribution (\ref{eq: boltzgibbs}), then the system (\ref{eq: general_system}) provides the maximum likelihood estimates for the parameters $\{ \theta_i \}$ \citep{PhysRevE.78.015101}.

When the constrained observables are the degree values $\{k_1, \dots, k_n\}$ of a binary symmetric network, the main quantities of the model read:

\begin{align*}
H(G) & =  \sum_i \sum_{j > i} \left[ (\theta_{i} + \theta_{j}) a_{ij} \right] = \sum_i \sum_{j > i} \Lambda_{ij} a_{ij}  \\
Z & =  \prod_i \prod_{j > i}  \left ( 1 + e^{-\Lambda_{ij}}\right ) \\
F & = - \ln Z = - \sum_i \sum_{j > i} \ln \left ( 1 + e^{-\Lambda_{ij}}\right )
\end{align*}

and $P(G)$ takes the form of the product of $n^2$ independent Bernoulli variables with parameters

\begin{equation} \label{eq: fermi_prob}
p_{ij} = \langle a_{ij} \rangle = \dfrac{\partial F}{\partial \Lambda_{ij}} = \dfrac{1}{e^{\Lambda_{ij}} + 1} \qquad \qquad i,j = 1, \dots, n
\end{equation}

Substituting the last equation into the constraints we obtain the following specialization of system (\ref{eq: general_system}):

\begin{equation} \label{eq: fermi_system}
\sum_{j \neq i}   \dfrac{1}{e^{\Lambda_{ij}} + 1} = \bar{k}_i \quad \quad i = 1, \dots n
\end{equation}

This system can be solved numerically in order to obtain the values ${\theta_i}$ which satisfy the constraints.

The corresponding system for weighted networks with given average strength distribution reads as follows:

\begin{equation} \label{eq: bose_system}
\sum_{j \neq i}   \dfrac{1}{e^{\Lambda_{ij}} - 1} = \bar{w}_i \quad \quad i = 1, \dots n
\end{equation}

In this case we obtain that the $w_{ij}$ are geometrically distributed. Unfortunately this ensemble has two main drawbacks: 1) the system (\ref{eq: bose_system}) is hard to solve; 2) even if a solution is obtained, the topology of networks in this ensemble is not bound to follow any topological property, such as the degree distribution or even connectivity. To put it short, the topology of this model is unrealistic by construction.

In order to overcome the first problem, a different maximum entropy technique is usually employed in the analysis of weighted networks, such as credit networks \citep{mistrulli2011assessing}. Following this approach, in the symmetric and weighted case, we wish to solve the following problem \footnote{The extension to the directed case is straightforward.}:

\begin{equation}\label{eq: max_ent_prob}
\max_{W} g\left( W \right)= - \sum_{i=1}^{n} \sum_{j=1}^{n} w_{ij} \ln w_{ij}
\end{equation}

subject to the following constraints:

\begin{equation}
\sum_{j=1}^{n} w_{ij} = \bar{w}_{i} \qquad \qquad i = 1 \dots n \label{foc1} \\
\end{equation}

where $\mathbf{\bar{w}} = \{\bar{w}_1, \dots,\bar{w}_n\}$ is a fixed strength sequence. It is well known that, when selfloops are allowed\footnote{When they are not, we can easily obtain a numerical solution starting from  the explicit formula mentioned in the text.}, the solution of this problem can be written down explicitly

\begin{equation} \label{eq: me_expect}
\bar{w}_{ij} = \frac{w_iw_j}{v}
\end{equation}

We see from problem (\ref{eq: max_ent_prob}) that this solution represents the most diversified configuration which is consistent with the given constraints. In the standard economic perspective, this property is convenient since microeconomic considerations dictate that the more diversification, the better for economic agents \citep{Allen2000}. In order to derive an ensemble for this configuration, we still need to assume a probability distribution for $w_{ij}$. Some common choices are the the Poisson distribution \citep{PhysRevE.83.016107}, or the binomial distribution \citep{bar2011396}.

The second problem is tackled rigorously in \cite{PhysRevLett.102.038701}, but the resulting models are in general difficult to solve numerically. For this reason here we adopt a less rigorous approach, and proceed to solve numerically the following system in the set of variables $\{x_1, \dots, x_n\}$:

\begin{equation} \label{eq: sparse_system}
\sum_{j \neq i} x_i x_j \dfrac{1}{e^{\Lambda_{ij}} + 1}  = \bar{w}_i \quad \quad i = 1, \dots n
\end{equation}

Here the $\Lambda_{ij}$ stand for the resolutive values of the system (\ref{eq: fermi_system}). Then we treat the weighted links of artificial networks in the ensemble as the product of two independent variables:

\begin{equation}
w_{ij} \sim \text{Bernoulli}(p_{ij}) \,\, \text{Poisson}(\lambda_{ij})
\end{equation}

with $p_{ij}$ obtained from (\ref{eq: fermi_system}) and $\lambda_{ij} = x_i x_j$ obtained from (\ref{eq: sparse_system}). It's easy to see that the ensemble obtained in this way satisfies simultaneously the constraints over degrees and strengths distributions.

    \bibliographystyle{chicago}

\end{document}